\documentclass[twocolumn]{pasj00}

\begin{document}
\SetRunningHead{Y.~Asaki et al.}
{Effectiveness of VSOP-2 Phase Referencing}
\Received{2006/07/03}
\Accepted{2006/12/25}

\title{
Verification of the Effectiveness of
VSOP-2 Phase Referencing with a Newly Developed Simulation Tool, ARIS
}

\author{
    Yoshiharu \textsc{Asaki},\altaffilmark{1,2}
    Hiroshi \textsc{Sudou},\altaffilmark{3}
    Yusuke \textsc{Kono},\altaffilmark{4}
    Akihiro \textsc{Doi},\altaffilmark{5}
    Richard \textsc{Dodson},\altaffilmark{6}
\\
    Nicolas \textsc{Pradel},\altaffilmark{4}
    Yasuhiro \textsc{Murata},\altaffilmark{1,2}
    Nanako \textsc{Mochizuki},\altaffilmark{1}
    Philip G. \textsc{Edwards},\altaffilmark{7}
\\
    Tetsuo \textsc{Sasao},\altaffilmark{8}
\\
    and
\\
    Edward B. {\sc Fomalont}\altaffilmark{9}
}
\altaffiltext{1}{
    Institute of Space and Astronautical Science, 
    3-1-1 Yoshinodai, Sagamihara, Kanagawa 229-8510
}
\altaffiltext{2}{
    Department of Space and Astronautical Science, 
    School of Physical Sciences, \\
    The Graduate University for Advanced Studies, 
    3-1-1 Yoshinodai, Sagamihara, Kanagawa 229-8510
}
\altaffiltext{3}{
    Department of Mathematical and Design Engineering, 
    Faculty of Engineering, Gifu University, \\
    1-1 Yanagido, Gifu, Gifu 501-1193
}
\altaffiltext{4}{
    National Astronomical Observatory, 
    2-21-1 Osawa, Mitaka, Tokyo 181-8588
}
\altaffiltext{5}{
    Department of Physics, Faculty of Science, Yamaguchi University, \\
    1677-1 Yoshida, Yamaguchi, Yamaguchi 753-8511
}
\altaffiltext{6}{
    Observatorio Astronomico Nacional, 
    Apartado 112 E-28803 Alcala de Henares Spain
}
\altaffiltext{7}{
    Australia Telescope National Facility, CSIRO, 
    Locked Bag 194 Narrabri NSW 2390 Australia
}
\altaffiltext{8}{
    Korea Astronomy and Space Science Institute, 
    61-1 Hwaam-dong, Yuseong-gu, Daejeon 305-348 Korea
}
\altaffiltext{9}{
    National Radio Astronomy Observatory, 520 Edgemont Road, 
    Charlottesville, VA 22903  USA
}
\email{
    asaki@vsop.isas.jaxa.jp,
    sudou@gifu-u.ac.jp,
    kono.yusuke@nao.ac.jp,
    doi@yamaguchi-u.ac.jp,
    r.dodson@oan.es,
\\
    nicolas.pradel@nao.ac.jp,
    murata@vsop.isas.jaxa.jp,
    nanakom@vsop.isas.jaxa.jp,
    philip.edwards@csiro.au,
\\
    sasao@ajou.ac.kr, 
    efomalon@nrao.edu
}

\KeyWords{instrumentation: high angular resolution---
space vehicles: instruments---
techniques: image processing}

\maketitle
\begin{abstract}
The next-generation space VLBI mission, VSOP-2, is expected to 
provide unprecedented spatial resolutions at 8.4, 22, and 43~GHz. 
In this report, phase referencing with VSOP-2 is examined in detail 
based on a simulation tool called ARIS. The criterion for 
successful phase referencing was to keep the phase errors below 
one radian. Simulations with ARIS reveal that phase referencing 
achieves good performance at 8.4~GHz, even under poor tropospheric 
conditions. At 22 and 43~GHz, it is recommended to conduct phase 
referencing observations under good or typical tropospheric 
conditions. The satellite is required to have an attitude-switching 
capability with a one-minute or shorter cycle, and an orbit 
determination accuracy higher than $\sim$10~cm at apogee; 
the phase referencing calibrators are required to have a signal-to-noise 
ratio larger than four for a single scan. The probability to find a 
suitable phase referencing calibrator was estimated by using 
VLBI surveys. From the viewpoint of calibrator availability, 
VSOP-2 phase referencing at 8.4~GHz is promising. However, the change 
of finding suitable calibrators at 22 and 43~GHz is significantly 
reduced; it is important to conduct specific investigations for 
each target at those frequencies. 
\end{abstract}
\section{
  Introduction
}

A series of space Very Long Baseline Interferometry (VLBI) 
experiments in which an orbiting radio telescope satellite 
was used for VLBI observations has provided successful 
development of very high spatial resolution in astronomy. 
In particular, the first dedicated space VLBI project, VSOP, with 
the HALCA satellite achieved remarkable scientific results based 
on spatial resolution up to 1.2 and 0.4~mas at 1.6 and 5~GHz, 
respectively, with an apogee altitude of 21375~km
\citep{Hirabayashi1998}. 
To come after the successful VSOP, the next space VLBI mission, 
VSOP-2, is being planned by Institute of Space and Astronautical 
Science (ISAS)
\citep{Hirabayashi2004}. 
This project will launch a satellite radio telescope (SRT), 
which will be equipped with a 9.1-m off-axis paraboloid 
antenna and dual-polarization receivers to observe at 8.4, 22, 
and 43~GHz, together with terrestrial radio telescopes (TRTs) 
with the sensitivity being a factor-of-ten higher than VSOP. 
Launch is planned for 2012. With two intermediate frequency (IF) 
bands, each with a two-bit sampled 128-MHz a total 256-MHz 
bandwidth will be available. 
The achievable maximum baselines will exceed
37800~km with the planned apogee altitude of 25000~km, 
and the highest spatial resolution will be 38~$\mu$as at 43~GHz. 

Millimeter-wave observations in space VLBI will be a frontier 
in astrophysics because there are various compact objects for 
which very high spatial resolution is essential 
\citep{Murphy2005}. 
There is, however, a difficulty in millimeter-wave VLBI 
in terms of the fringe phase stability. Data-averaging of 
the fringe is usually performed within a certain time scale, 
the so-called coherence time, for which a root-mean-square (RMS) 
of the fringe phase is less than one radian. In the conventional 
calibration scheme in VLBI with a fringe-fitting technique 
\citep{Cotton1995}, 
it is necessary that the fringe be detected in less than the 
coherence time. The coherence time in VLBI at 43~GHz 
is limited to a few minutes by stochastic 
variations in the fringe phase, mainly due to the turbulent 
media of the Earth's atmosphere; it is thus difficult to conduct 
a long-time averaging in VLBI in order to improve the 
signal-to-noise ratio (SNR). 
Although celestial radio waves received on a satellite are 
not affected by the atmosphere, the fringe of a space baseline 
(a combination of orbiting and terrestrial telescopes) 
also suffers from the atmospheric phase fluctuations because 
one of the elements is inevitably a terrestrial radio telescope. 

Phase referencing is a successful phase-calibration 
scheme for VLBI. Here a scientifically interesting 
target source is observed with an adjacent reference 
calibrator with fast antenna pointing changes 
(antenna switching) in order to compensate for any rapid 
phase fluctuations due to the atmosphere 
(\cite{Beasley1995}; 
\cite{Fomalont1995}). 
Phase referencing can also remove long-term phase 
drifts due to geometrical errors and smoothly variable 
atmospheric delay errors, as well as any instability of the 
independent frequency standards. 
The phase referencing technique has been proved for 
imaging faint radio sources that cannot be detected 
with the conventional VLBI data reduction
(\cite{Smith2003}; 
\cite{Gallimore2004a}; 
\cite{Gallimore2004b}). 
Various astrometric observations in the VLBI field have 
also been made with the phase referencing technique 
to obtain the relative positions with the accuracies
on the order of 10~$\mu$as 
(\cite{Shapiro1979}; 
\cite{Bartel1986}; 
\cite{Gwinn1986}; 
\cite{Lestrade1990}; 
Reid et al. 1999;
\cite{Brunthaler2005}).

Phase referencing will be useful for VSOP-2. To investigate which 
component of errors has a more significant influence than others 
on the quality of the synthesized images obtained with 
VSOP-2 phase referencing, and to give feedback for designing 
the satellite system, a software simulation tool for space VLBI 
has been developed. Here we report on the effectiveness of VSOP-2 
phase referencing by simulation. 
The basic ideas of phase compensation with phase referencing 
are described in section~2. The residual phase errors 
after phase compensation are discussed in section~3 based on 
quantitative estimations. In section~4, the newly developed 
space VLBI simulator for this study is described. Simulation work 
for VSOP-2 phase referencing is presented to clarify the constraints 
on specific observing parameters: the separation angle between 
a target and a calibrator; the time interval to switch 
a pair of sources; the orbit determination accuracy of the 
satellite. Further discussions are given in section~5 to describe 
the feasibility of phase referencing with VSOP-2. The conclusions 
are summarized in section~6. 

\section{
  Basic Ideas of VLBI Phase Referencing
}

\begin{figure}
  \begin{center}
    \FigureFile(80mm,80mm){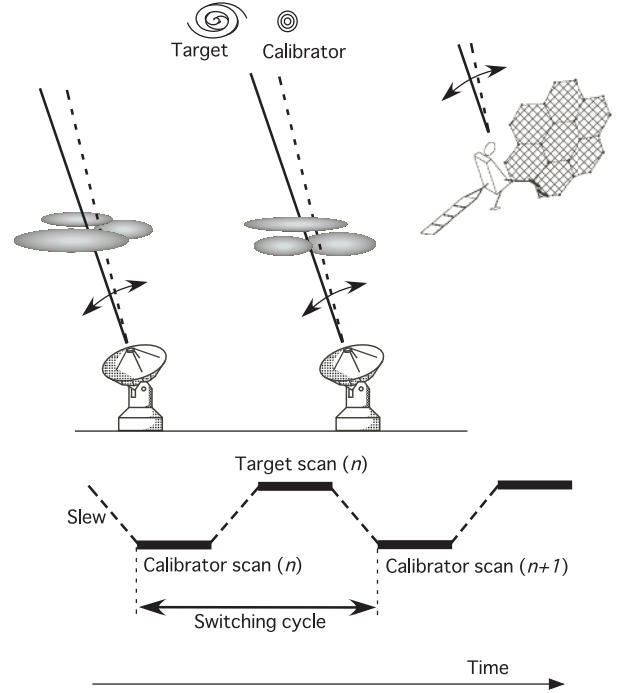}
  \end{center}
  \caption{
Schematic drawing of VSOP-2 phase referencing. 
  }\label{fig:02-01}
\end{figure}

In this section we describe basic ideas of phase referencing. 
In the following discussion a single on-source duration 
is referred to as a scan. In phase referencing, alternate 
scans are made on the target and phase referencing calibrator. 
One observation period from the beginning of the calibrator 
scan, then the target scan and return to the beginning of 
the calibrator scan, is referred to as the switching cycle time, 
$T_{\mathrm{swt}}$. 
Figure~\ref{fig:02-01} shows a schematic drawing of 
VSOP-2 phase referencing. 
The correlated VLBI data reveal a time series of a complex 
quantity, called a fringe, which is composed of amplitude 
and phase including information on the visibility of a celestial 
object as well as various errors from instruments and propagation 
media. Let us assume that the difference in the arrival time of 
a celestial radio wave between 
telescopes and its derivative are largely removed by subtracting 
their a priori values, calculated in the correlator. 

Phase referencing is used to observe the target and a closely 
located calibrator in the sky. We refer to the fringe phases 
of the target and phase referencing calibrator as 
$\Phi^{\mathrm{t}}$ and $\Phi^{\mathrm{c}}$, 
respectively, expressed as follows: 
\begin{eqnarray}
\label{equ:02-01}
\Phi^{\mathrm{t}}(t^{\mathrm{t}})
 - \Phi^{\mathrm{t}}_{\mathrm{apri}}(t^{\mathrm{t}}) &=&
   \Phi^{\mathrm{t}}_{\mathrm{dtrp}}(t^{\mathrm{t}})
 + \Phi^{\mathrm{t}}_{\mathrm{dion}}(t^{\mathrm{t}})
 + \Phi^{\mathrm{t}}_{\mathrm{strp}}(t^{\mathrm{t}}) 
  \nonumber \\
          &+&
   \Phi^{\mathrm{t}}_{\mathrm{sion}}(t^{\mathrm{t}})
 + \Phi^{\mathrm{t}}_{\mathrm{bl}}(t^{\mathrm{t}})
 + \Phi^{\mathrm{t}}_{\mathrm{inst}}(t^{\mathrm{t}})
  \nonumber \\
          &+&
   \Phi^{\mathrm{t}}_{\mathrm{\Delta s}}(t^{\mathrm{t}})
 + \Phi^{\mathrm{t}}_{\mathrm{v}}(t^{\mathrm{t}})
 + \epsilon^{\mathrm{t}}_{\mathrm{therm}}(t^{\mathrm{t}}), \\
\label{equ:02-02}
\Phi^{\mathrm{c}}(t^{\mathrm{c}})
 - \Phi^{\mathrm{c}}_{\mathrm{pred}}(t^{\mathrm{c}}) &=&
   \Phi^{\mathrm{c}}_{\mathrm{dtrp}}(t^{\mathrm{c}})
 + \Phi^{\mathrm{c}}_{\mathrm{dion}}(t^{\mathrm{c}})
 + \Phi^{\mathrm{c}}_{\mathrm{strp}}(t^{\mathrm{c}})
  \nonumber \\
          &+&
   \Phi^{\mathrm{c}}_{\mathrm{sion}}(t^{\mathrm{c}})
 + \Phi^{\mathrm{c}}_{\mathrm{bl}}(t^{\mathrm{c}})
 + \Phi^{\mathrm{c}}_{\mathrm{inst}}(t^{\mathrm{c}})
  \nonumber \\
          &+&
   \Phi^{\mathrm{c}}_{\mathrm{\Delta s}}(t^{\mathrm{c}})
 + \Phi^{\mathrm{c}}_{\mathrm{v}}(t^{\mathrm{c}})
 + \epsilon^{\mathrm{c}}_{\mathrm{therm}}(t^{\mathrm{c}}),
\end{eqnarray}
where 
\begin{list}{}
{
  \setlength{\itemindent}{0mm}
  \setlength{\parsep}{0mm}
  \setlength{\topsep}{0mm}
}
\item{$\Phi_{\mathrm{apri}}$ : }
the a priori phase calculated in the correlator; 

\item{$\Phi_{\mathrm{dtrp}}$, $\Phi_{\mathrm{dion}}$ : }
phase errors due to the dynamic components 
of the troposphere and ionosphere, respectively; 

\item{$\Phi_{\mathrm{strp}}$, $\Phi_{\mathrm{sion}}$ : }
long-term phase variations depending on the observing 
elevations of terrestrial telescopes due to the uncertainties 
of the tropospheric and ionospheric zenith excess path delays, 
respectively; 

\item{$\Phi_{\mathrm{bl}}$ : }
phase error due to the baseline vector error coming from 
uncertainties of the telescope positions and erroneous estimations 
of the Earth Orientation Parameters (EOP); 

\item{$\Phi_{\mathrm{inst}}$ : }
the instrumental phase error due to the independent frequency 
standards, transmitting electric cables and so on; 

\item{$\Phi_{\mathrm{\Delta s}}$ : } 
phase error due to the uncertainty in 
the a priori source position in the sky; 

\item{$\Phi_{\mathrm{v}}$ : } 
visibility phase component representing the source structure; 

\item{$\epsilon_{\mathrm{therm}}$ : } 
a contribution of the thermal noise. 

\end{list}{}
Here, $t^{\mathrm{t}}$ is the time that the target is observed, and 
$t^{\mathrm{c}}$, temporally $T_{\mathrm{swt}}/2$ apart from 
$t^{\mathrm{t}}$, is the time that the calibrator is observed. 

If the structure of the calibrator is well-known, 
three terms can be identified in equation~(\ref{equ:02-02}), 
as follows: 
\begin{eqnarray}
\label{equ:02-03}
\Phi^{\mathrm{c}}(t^{\mathrm{c}})
 - \Phi^{\mathrm{c}}_{\mathrm{apri}}(t^{\mathrm{c}}) &=&
   \Phi^{\mathrm{c}}_{\mathrm{err}}(t^{\mathrm{c}})
 + \Phi^{\mathrm{c}}_{\mathrm{v}}(t^{\mathrm{c}}),
\end{eqnarray}
where 
$\Phi^{\mathrm{c}}_{\mathrm{err}}$ 
is an error term consisting of 
$\Phi^{\mathrm{c}}_{\mathrm{dtrp}}$, 
$\Phi^{\mathrm{c}}_{\mathrm{dion}}$, 
$\Phi^{\mathrm{c}}_{\mathrm{strp}}$, 
$\Phi^{\mathrm{c}}_{\mathrm{sion}}$, 
$\Phi^{\mathrm{c}}_{\mathrm{bl}}$, 
$\Phi^{\mathrm{c}}_{\mathrm{inst}}$, 
$\Phi^{\mathrm{c}}_{\mathrm{\Delta s}}$, 
and 
$\epsilon^{\mathrm{c}}_{\mathrm{therm}}$. 
The calibration data, $\Phi_{\mathrm{cal}}$, for the target at 
time $t^{\mathrm{t}}$ is obtained from 
$\Phi^{\mathrm{c}}_{\mathrm{err}}$ 
of the temporally closest two calibrator scans, as follows: 
\begin{eqnarray}
\label{equ:02-04}
\Phi_{\mathrm{cal}}(t^{\mathrm{t}}) &=&
  \frac{
   \Phi^{\mathrm{c}}_{\mathrm{err}}(t^{\mathrm{t}}-T_{\mathrm{swt}}/2)
 + \Phi^{\mathrm{c}}_{\mathrm{err}}(t^{\mathrm{t}}+T_{\mathrm{swt}}/2)}{2}
  \nonumber \\
              &=&
   \Phi^{'\mathrm{c}}_{\mathrm{dtrp}}(t^{\mathrm{t}})
 + \Phi^{'\mathrm{c}}_{\mathrm{dion}}(t^{\mathrm{t}})
 + \Phi^{'\mathrm{c}}_{\mathrm{strp}}(t^{\mathrm{t}})
  \nonumber \\
              & &+
   \Phi^{'\mathrm{c}}_{\mathrm{sion}}(t^{\mathrm{t}})
 + \Phi^{'\mathrm{c}}_{\mathrm{bl}}(t^{\mathrm{t}})
 + \Phi^{'\mathrm{c}}_{\mathrm{inst}}(t^{\mathrm{t}}) 
  \nonumber \\
              & &+
   \Phi^{'\mathrm{c}}_{\mathrm{\Delta s}}(t^{\mathrm{t}})
 + \epsilon^{'\mathrm{c}}_{\mathrm{therm}}(t^{\mathrm{t}}),
\end{eqnarray}
where 
$\Phi^{'\mathrm{c}}(t^{\mathrm{t}})$ 
and 
$\epsilon^{'\mathrm{c}}_{\mathrm{therm}}$ 
are interpolated calibrator phases at $t^{\mathrm{t}}$. 
The final step of the phase compensation is carried out by subtracting 
$\Phi_{\mathrm{cal}}(t^{\mathrm{t}})$ from equation (\ref{equ:02-01}), 
as follows: 
\begin{eqnarray}
\label{equ:02-05}
&&\Phi^{\mathrm{t}}(t^{\mathrm{t}})
 - \Phi^{\mathrm{t}}_{\mathrm{apri}}(t^{\mathrm{t}})
 - \Phi_{\mathrm{cal}}(t^{\mathrm{t}})
  \nonumber \\
                          &=&
   \Phi^{\mathrm{t}}_{\mathrm{v}}(t^{\mathrm{t}})
 + \left [
   \Phi^{\mathrm{t}}_{\mathrm{\Delta s}}(t^{\mathrm{t}})
 - \Phi^{'\mathrm{c}}_{\mathrm{\Delta s}}(t^{\mathrm{t}})
   \right ]
  \nonumber \\
                          & &+
   \phi_{\mathrm{dtrp}}(t^{\mathrm{t}})
 + \phi_{\mathrm{dion}}(t^{\mathrm{t}})
 + \phi_{\mathrm{strp}}(t^{\mathrm{t}})
 + \phi_{\mathrm{sion}}(t^{\mathrm{t}})
  \nonumber \\
                          & &+
   \phi_{\mathrm{bl}}(t^{\mathrm{t}})
 + \phi_{\mathrm{inst}}(t^{\mathrm{t}})
 + \Delta\epsilon_{\mathrm{therm}}(t^{\mathrm{t}}),
\end{eqnarray}
where $\phi$ is the phase difference between 
$\Phi^{\mathrm{t}}$ 
and 
$\Phi^{'\mathrm{c}}$, 
and $\Delta\epsilon_{\mathrm{therm}}$ is the phase difference between 
$\epsilon^{\mathrm{t}}_{\mathrm{therm}}$ 
and 
$\epsilon^{'\mathrm{c}}_{\mathrm{therm}}$. 
If $T_{\mathrm{swt}}$ is short enough (typically, shorter than a 
few minutes at centimeter to millimeter waves) and the calibrator 
is located closely enough (typically, within a few degrees), 
the phase errors, except for the thermal noise, can almost be canceled. 
An uncanceled term, 
$
   \Phi^{\mathrm{t}}_{\mathrm{\Delta s}}(t^{\mathrm{t}})
 - \Phi^{'\mathrm{c}}_{\mathrm{\Delta s}}(t^{\mathrm{t}})
$, 
gives the relative position of the target to the calibrator 
with a typical accuracy of much less than one mas. 
Another aspect of the advantages of phase referencing is 
to eliminate the rapid time variation caused by the turbulent 
atmosphere. This means that the coherence time, which is limited 
by the atmosphere, becomes longer, so that faint radio sources 
can be detected by means of the long time averaging. 

\section{
  Residual Phase Errors after Phase Referencing with VSOP-2
}

Although phase referencing is capable of removing 
a large amount of the fringe phase errors, residual 
phase errors remain after phase compensation 
because the target and calibrator are observed with a 
certain time separation, and not on the same line of sight. 
Being different from a terrestrial baseline consisting 
of both terrestrial telescopes, a single space baseline 
includes a single line-of-sight atmospheric phase error 
for a terrestrial telescope, and uncertainty of a satellite 
trajectory in the orbit. In this section we analytically 
estimate the residual phase errors of the space baseline 
after phase compensation. We attempt 
to characterize the distribution of phase offsets from 
a very large number of samples obtained with given 
parameters, such as uncertainties in the 
a priori values, switching cycle time, separation angle, 
zenith angle, and so on. 

\subsection{
  Fringe Phase Fluctuations due to a Turbulent Atmosphere
}

The Earth's atmosphere causes an excess path of radio waves 
passing through it 
(Thompson et al. 2001). 
Let us distinguish two types of excess path errors in the 
following discussions: one is a dynamic component 
(fluctuation error), and the other one is a nearly static 
component (systematic error). 
Here, we first address the atmospheric phase fluctuations. 

\subsubsection{
  Statistical model of atmospheric phase fluctuations
}

Phase differences due to a turbulent medium are often 
characterized by a spatial structure function (SSF), 
which is defined as the mean-square difference in the 
phase fluctuations at two sites separated by a displacement 
vector, $\mbox{\boldmath $\rho$}$, 
\citep{Tatarskii1961} 
as follows: 
\begin{eqnarray}
\label{equ:03-01}
D(\rho) &=& \left< \left[ 
              \Phi(\mbox{\boldmath $x$} + \mbox{\boldmath $\rho$})
            - \Phi(\mbox{\boldmath $x$})
                   \right]^2
            \right>, 
\end{eqnarray}
where the angle brackets mean an ensemble average, and 
$\mbox{\boldmath $x$}$ is a position vector. 
According to 
\citet{Dravskikh1979}, 
the SSF of the atmospheric phase shift with Kolmogorov turbulence 
can be approximated by: 
\begin{eqnarray}
\label{equ:03-02}
D(\rho) &=&
  C^2 \rho^{\frac{5}{3}}
\ \ \ \ \ \ \ \ \ \ \ (\rho \leq L_1), 
\\
\label{equ:03-03}
D(\rho) &=&
  C^2 L_1 \rho^{\frac{2}{3}}
\ \ \ \ \ \ \ (L_1 < \rho \leq L_2), 
\\
\label{equ:03-04}
D(\rho) &=&
  C^2 L_1 L_2^{\frac{2}{3}}
\ \ \ \ \ \ \ (\rho > L_2),
\end{eqnarray}
where $L_1$ and $L_2$ are referred as inner and outer scales, 
respectively, and $C$ is a structure coefficient. 

The spatial structure of the atmospheric phase fluctuations 
can also be modeled as a phase screen, assuming frozen flow 
\citep{Dravskikh1979}, 
in which a laminar sheet with a fixed spatial pattern 
representing the distribution of the turbulent medium, 
flows at a constant speed. 
We use the phase screen model in the following discussions 
about the atmospheric phase fluctuations. 

\subsubsection{
  Tropospheric phase fluctuations
}

The water vapor in the troposphere is neither well mixed 
nor regularly distributed in the lower troposphere, and 
therefore is not well correlated with the ground-based 
meteorological parameters. Hence, the water vapor in the 
troposphere is highly unpredictable. 
\citet{Carilli1999} 
observationally showed that the inner and outer scales of 
the SSF are 1.2 and 6~km, respectively, at their observing 
site at the Very Large Array. 
The tropospheric phase fluctuations play a major role in 
restricting the coherence time to be quite short, especially 
at millimeter waves, because the refractive index is non-dispersive 
in centimeter-to-millimeter waves. At higher observing frequency 
bands, VLBI synthesized images can be improved 
by removing telescope data whose weather is not good. 
Thus, to use a large terrestrial telescope network 
in VLBI is important not only for ($u$,~$v$) coverage, but 
also to have a number of telescopes where the tropospheric 
conditions are good. 

To make a successful phase connection at 22 or 43~GHz by 
interpolating the calibrator fringe phases, the switching 
cycle time should be shorter than a few minutes, and the 
separation angle should be smaller than a few degrees. 
Assuming such a fast antenna switching observations for such a 
pair of sources, a residual phase error, 
$\sigma_{\phi_{\mathrm{dtrp}}}$, 
due to the dynamic troposphere for a space baseline, is 
expressed as follows:   
\begin{eqnarray}
\label{equ:03-05}
\sigma_{\phi_{\mathrm{dtrp}}}~[\mathrm{rad}]
          &\approx&
                    \frac{2\pi\nu C_{\mathrm{n}}
                    \sqrt{1.4H_{\mathrm{s}}}
                    \sqrt{\sec{Z_{\mathrm{g}}}}}{c}
          \nonumber \\
          &\times&
                  \left(\frac{v_{\mathrm{w}} T_{\mathrm{swt}}}{2}
                   +{H_{\mathrm{w}} \Delta\theta}\sec{Z_{\mathrm{g}}}
                  \right)^\frac{5}{6},
\end{eqnarray}
where 
$\Delta\theta$ is the separation angle between the target 
and the calibrator, 
$\nu$ is the observing frequency, 
$c$ is the speed of light, 
$C_{\mathrm{n}}$ is the structure coefficient of the SSF 
of the troposphere to the zenith, defined by 
\citet{Beasley1995}, 
$H_{\mathrm{s}}$ is the scale height of the tropospheric water vapor, 
$v_{\mathrm{w}}$ is the wind velocity aloft or flow speed 
of the phase screen, 
$H_{\mathrm{w}}$ is the height of the phase screen, and 
$Z_{\mathrm{g}}$ is the zenith angle at the ground surface for 
a terrestrial telescope. 
The first $\sec{Z_{\mathrm{g}}}$ in equation (\ref{equ:03-05}) 
comes from an indication based on numerical calculations for 
the tropospheric phase fluctuations made by 
\citet{Treuhaft1987}, 
while the second comes from a factor to project the 
separation between a pair of sources onto the phase screen. 
Although $C_{\mathrm{n}}$ is very different at different 
telescope sites, seasons, and weather conditions, 
it can be assumed that the values of 
1$\times 10^{-7}$, 
2$\times 10^{-7}$, and 
4$\times 10^{-7}$~m$^{-1/3}$ 
are equivalent to good, typical, and poor tropospheric conditions, 
respectively, with the assumption of Kolmogorov turbulence 
(J. Ulvestad, VLBA Scientific Memo No.20).\footnote{
$<$http://www.vlba.nrao.edu/memos/sci/.$>$}
Assuming typical values of 
$H_{\mathrm{s}}=H_{\mathrm{w}}=$1~km and $v_{\mathrm{w}}=$10~m~s$^{-1}$, 
we can obtain the following approximation from equation (\ref{equ:03-05}): 
\begin{eqnarray}
\label{equ:03-07}
&&\sigma_{\phi_{\mathrm{dtrp}}}~\mathrm{[deg]}
            \approx
            27 C_{\mathrm{w}} \cdot
                  \left(
                      \frac{\nu~\mathrm{[GHz]}}{43~\mathrm{GHz}}
                  \right)
                  \left(
                      \frac{\sec{Z_{\mathrm{g}}}}{\sec{45^\circ}}
                  \right)^\frac{1}{2}
                                     \nonumber \\
              &\times&
                  \left[
                  \left(
                      \frac{T_{\mathrm{swt}}~\mathrm{[s]}}{60~\mathrm{s}}
                  \right)
                       + 0.16 \cdot 
                  \left(
                      \frac{\sec{Z_{\mathrm{g}}}}{\sec{45^\circ}}
                  \right)
                  \left(
                         \frac{\Delta\theta~\mathrm{[deg]}}{2^\circ}
                  \right)
                  \right]^\frac{5}{6},
\end{eqnarray}
where $C_{\mathrm{w}}$ is a modified structure coefficient of 
the SSF, whose values are 1, 2, and 4 for good, typical, and 
poor tropospheric conditions, respectively. 
The residual phase errors with the typical parameters used in 
equation (\ref{equ:03-07}) and $C_{\mathrm{w}}=2$ are given 
in table~\ref{tbl:03-01}. 

\subsubsection{
  Ionospheric phase fluctuations
}

Ionospheric phase fluctuations are caused by irregularities 
of the plasma density in the ionosphere. 
Since the extra phase due to the Total Electron Content (TEC) 
has an inverse proportionality to radio frequency, the amplitude 
of the ionospheric phase fluctuations becomes smaller as 
the observing frequency increases. 

In this report we focus on the 
temporal TEC variations known as medium-scale traveling 
ionospheric disturbances (MS-TIDs), which have a severe 
influence on the VLBI observables, especially at frequencies 
less than 10~GHz. MS-TIDs, firstly classified by 
\citet{Georges1968} 
are often seen at night in high- and mid-latitude areas, 
and thought to be caused by the thermospheric gravity sound 
waves at the bottom of the F-region. 
Studies of spectra of the gravity waves in the high-latitude 
thermosphere by 
\citet{Bristow1997} 
showed that monochromatic waves with a period of a few tens 
of minutes are present. In addition, the power spectra, 
ranging from 0.3 to a few milli hertz, show characteristics 
of Kolmogorov power-law, which indicates that kinematic energy 
causing the power-law perturbations are injected from the 
monochromatic gravity sound waves. Typical MS-TIDs in 
mid-latitudes have a spatial wavelength of a few hundred 
kilometers and a propagation velocity of around 100 ms$^{-1}$ 
from high- to low-latitudes 
\citep{Saito1998}.

We here attempt to make a model for ionospheric phase fluctuations 
with the assumption of a phase screen with Kolmogorov turbulence 
driven by the MS-TIDs. The residual phase error, 
$\sigma_{\phi_{\mathrm{dion}}}$, 
due to the dynamic ionosphere for a space baseline, 
is expressed as follows:   
\begin{eqnarray}
\label{equ:03-08}
\sigma_{\phi_{\mathrm{dion}}}~[\mathrm{rad}]
          &\approx&
                    \frac{2\pi \kappa C_{\mathrm{i}}
                    \sqrt{\sec{Z_{\mathrm{i}}}}}{c\nu}
          \nonumber \\
          &\times&
              \left(\frac{v_{\mathrm{i}} T_{\mathrm{swt}}}{2}
                 +H_{\mathrm{i}}\Delta\theta\sec{Z_{\mathrm{i}}}
                  \right)^\frac{5}{6}, 
\end{eqnarray}
where 
$\kappa=40.3$~m$^3$s$^{-2}$, 
$C_{\mathrm{i}}$ is the structure coefficient of the SSF of the 
ionospheric TEC to the zenith, 
$v_{\mathrm{i}}$ is the flow speed of the screen, 
and $H_{\mathrm{i}}$ is the phase screen height. 
$Z_{\mathrm{i}}$ is the zenith angle of the ray at 
$H_{\mathrm{i}}$, as follows: 
\begin{eqnarray}
\label{equ:03-09}
Z_{\mathrm{i}} & = &
    \arcsin{
      \left(
        \frac{R_{\oplus}}{R_{\oplus} + H_{\mathrm{i}}}
        \sin{Z_{\mathrm{g}}}
      \right)
    }, 
\end{eqnarray}
where $R_{\oplus}$ is the Earth radius. 

As described in subsubsection \ref{TEC_DATA_ANALYSES}, 
the value of $C_{\mathrm{i}}$ for a 50-percentile condition is 
roughly 
$1.6\times10^{-5}$~TECU$\cdot$m$^{-5/6}$ at mid-latitudes, 
where TECU is the unit used for TEC (1~TECU$=10^{16}$~electrons~m$^{-2}$). 
We can obtain the following approximation from equation (\ref{equ:03-08}) 
for the 50-percentile condition with assumptions of 
$v_{\mathrm{i}}=100$~ms$^{-1}$ and 
$H_{\mathrm{i}}=300$~km (bottom of the F-region): 
\begin{eqnarray}
\label{equ:03-10}
&&\sigma_{\phi_{\mathrm{dion}}}~\mathrm{[deg]}
            \approx
            0.46 \cdot
                  \left(
                      \frac{\sec{Z_{\mathrm{i}}}}{\sec{43^\circ}}
                  \right)^{\frac{1}{2}}
                  \left(
                      \frac{\nu~\mathrm{[GHz]}}{43~\mathrm{GHz}}
                  \right)^{-1}
                             \nonumber \\
           &\times&
                  \left[
                  0.21 \cdot
                  \left(
                      \frac{T_{\mathrm{swt}}~\mathrm{[s]}}{60~\mathrm{s}}
                  \right)
                       + 
                  \left(
                      \frac{\sec{Z_{\mathrm{i}}}}{\sec{43^\circ}}
                  \right)
                  \left(
                      \frac{\Delta\theta~\mathrm{[deg]}}{2^\circ}
                  \right)
                  \right]^\frac{5}{6}.
\end{eqnarray}
Note that $Z_{\mathrm{i}}$ of $43^\circ$ is the zenith angle at 
$H_{\mathrm{i}}=300$~km when $Z_{\mathrm{g}}$ is $45^\circ$. 
The residual phase errors with the typical parameters used 
in equation (\ref{equ:03-10}) are shown in table~\ref{tbl:03-01}. 

\subsection{
  Static Components of the Troposphere and Ionosphere
}

Let us focus on the atmospheric excess path errors after subtracting 
the dynamic components, which we assume to be temporally stable 
during an observation for up to several hours. The systematic 
errors are mainly caused by the uncertainty of the tropospheric 
water vapor constituent, and an inaccurate estimate of vertical 
TEC (VTEC). Hereafter, the elevation dependence of the atmospheric 
line-of-sight excess path (mapping function) is approximated by 
$\sec{Z}$ at the height of a homogeneously distributed medium. 

The residual phase error, $\sigma_{\phi_{\mathrm{strp}}}$, 
for a space baseline, due to an inaccurate estimate of 
the tropospheric zenith excess path, is expressed as 
\begin{eqnarray}
\label{equ:03-11}
\sigma_{\phi_{\mathrm{strp}}}~\mathrm{[rad]}
      &\approx&
      \frac{2\pi\nu \Delta l_{\mathrm{z}}
        \Delta Z_{\mathrm{g}}\sec{Z_{\mathrm{g}}}\tan{Z_{\mathrm{g}}}}{c},
\end{eqnarray}
where 
$\Delta l_{\mathrm{z}}$ is the tropospheric systematic error 
of the excess path length to the zenith, and 
$\Delta Z_{\mathrm{g}}$ is a difference of the zenith angles 
between the target and the calibrator. 
Assuming a $\Delta l_{\mathrm{z}}$ of 3~cm (Reid et al. 1999) 
and $\Delta Z_{\mathrm{g}}\sim\Delta\theta$, the following 
approximation can be obtained from equation~(\ref{equ:03-11}): 
\begin{eqnarray}
\label{equ:03-12}
\sigma_{\phi_{\mathrm{strp}}}&&~[\mathrm{deg}]
       \approx
       76 \cdot 
          \left(
            \frac{\nu~\mathrm{[GHz]}}{43~\mathrm{GHz}}
          \right)
          \left(
            \frac{\Delta l_{\mathrm{z}}~\mathrm{[cm]}}{3~\mathrm{cm}}
          \right)
    \nonumber \\
    &&\times
          \left(
            \frac{\Delta\theta~\mathrm{[deg]}}{2^\circ}
          \right)
          \left(
            \frac{\cos{Z_{\mathrm{g}}}}{\cos{45^\circ}}
          \right)^{-1} 
          \left(
            \frac{\tan{Z_{\mathrm{g}}}}{\tan{45^\circ}}
          \right). 
\end{eqnarray}
The residual phase errors with the typical parameters used 
in equation (\ref{equ:03-12}) are shown in table~\ref{tbl:03-01}. 

A residual phase error $\sigma_{\phi_{\mathrm{sion}}}$ 
for a space baseline due to an inaccurate TEC measurement 
is expressed as follows: 
\begin{eqnarray}
\label{equ:03-13}
\sigma_{\phi_{\mathrm{sion}}} \mathrm{[rad]}
      &\approx&
    \frac{2\pi\kappa
        \Delta I_{\mathrm{v}} \Delta Z_{\mathrm{F}}
        \sec{Z_{\mathrm{F}}}\tan{Z_{\mathrm{F}}}}{c\nu},
\end{eqnarray}
where $\Delta I_{\mathrm{v}}$ is the VTEC systematic error, 
$Z_{\mathrm{F}}$ is the zenith angle at the altitude of the 
electron density peak (typically, 450~km), and 
$\Delta Z_{\mathrm{F}}$ is the difference of the zenith angles 
between the target and the calibrator. 
A TEC measurement technique with the Global Positioning 
System (GPS) has been used to correct for any line-of-sight 
excess path delays due to the ionosphere 
(e.g., C.~Walker \& S.~Chatterjee 1999, VLBA Scientific Memo 
No.23;\footnote{
$<$http://www.vlba.nrao.edu/memos/sci/$>$.} 
\cite{Ros2000}). 
For example, global ionospheric modeling with GPS generally 
has an accuracy of 3--10~TECU, or at 10--20\% level 
\citep{Ho1997}. 
Let us assume $\Delta Z_{\mathrm{F}}\sim\Delta\theta$ and 
$Z_{\mathrm{g}}$ of $45^\circ$ to obtain the following 
approximation from equation~(\ref{equ:03-13}): 
\begin{eqnarray}
\label{equ:03-14}
&&\sigma_{\phi_{\mathrm{sion}}}~\mathrm{[deg]}
       \approx
       2.7 \cdot 
          \left(
            \frac{\nu~\mathrm{[GHz]}}{43~\mathrm{GHz}}
          \right)^{-1}
                         \nonumber \\
                    &\times&
          \left(
            \frac{\Delta I_{\mathrm{v}}~\mathrm{[TECU]}}{6~\mathrm{TECU}}
          \right)
                         \nonumber \\
                    &\times&
          \left(
            \frac{\Delta\theta~\mathrm{[deg]}}{2^\circ}
          \right)
          \left(
            \frac{\cos{Z_{\mathrm{F}}}}{\cos{41^\circ}}
          \right)^{-1}
          \left(
            \frac{\tan{Z_{\mathrm{F}}}}{\tan{41^\circ}}
          \right). 
\end{eqnarray}
The residual phase errors with the typical parameters used 
in equation~(\ref{equ:03-14}) are given in table~\ref{tbl:03-01}. 

\subsection{
  Baseline Errors
}

One of the special issues related to the space VLBI is the 
satellite orbit determination (OD) error. In VSOP, the precisely 
reconstructed OD of HALCA has an accuracy of 2--5~m 
\citep{Porcas2000} 
with the  $S$-band range and range-rate, and the $Ku$-band Doppler 
measurements, which is the best accuracy achieved by 
the Doppler tracking. However, the typical accuracy of terrestrial 
telescope positions is around 1~cm. For VSOP-2, as discussed in 
subsection \ref{OD_METHOD}, cm-order OD accuracy will be achieved 
by using an on-board GPS receiver. 

The accuracies of the EOP solutions given by International 
Earth Rotation and Reference Systems Service (IERS) are typically 
0.1~mas in the terrestrial pole offset, 0.3~mas in the celestial 
pole offset, and 0.02~ms in the UT1 offset.  
These uncertainties may cause the additional displacement of 
a few cm for the space baseline. 
A residual phase error, $\sigma_{\phi_{\mathrm{bl}}}$, 
for a space baseline due to the baseline error 
is approximated as follows: 
\begin{eqnarray}
\label{equ:03-15}
\sigma_{\phi_{\mathrm{bl}}}&&~\mathrm{[rad]}
       \approx
      \nonumber \\
 && \frac{\sqrt{2}\pi\nu\Delta\theta}{c}
          \sqrt{\Delta P_{\mathrm{TRT}}^2
              + \Delta P_{\mathrm{SRT}}^2 + B^2\Delta\Theta^2},
\end{eqnarray}
where $\Delta P_{\mathrm{TRT}}$ is the uncertainty of a 
terrestrial telescope position adopted in the correlator, 
$\Delta P_{\mathrm{SRT}}$ is a displacement of the OD error 
of a satellite radio telescope, 
$B$ is the projected baseline length to the celestial sphere 
at the source, and  
$\Delta\Theta$ is the EOP error. 
Equation~(\ref{equ:03-15}) can be expressed by the following 
approximation: 
\begin{eqnarray}
\label{equ:03-16}
\sigma_{\phi_{\mathrm{bl}}}&&~\mathrm{[deg]}
      \ \ \approx\ \  
          13 \cdot 
          \left(
            \frac{\nu~\mathrm{[GHz]}}{43~\mathrm{GHz}}
          \right)
          \left(
            \frac{\Delta\theta~\mathrm{[deg]}}{2^\circ}
          \right)
                     \nonumber \\
      &&\times
          \left[
            \left(
               \frac{\Delta P_{\mathrm{TRT}}~\mathrm{[cm]}}{1~\mathrm{cm}}
            \right)^2
         +  \left(
               \frac{\Delta P_{\mathrm{SRT}}~\mathrm{[cm]}}{1~\mathrm{cm}}
            \right)^2
          \right .
        \nonumber \\
        &&+
          \left .
                5.9\cdot
                \left(
                   \frac{B~\mathrm{[km]}}{25,000~\mathrm{km}}
                \right)^2
                \left(
                   \frac{\Delta\Theta~\mathrm{[mas]}}{0.2~\mathrm{mas}}
                \right)^2
           \right]^{\frac{1}{2}}.
\end{eqnarray}
The residual phase errors with the typical parameters used 
in equation~(\ref{equ:03-16}) are given in table~\ref{tbl:03-01}. 

\subsection{
  Instrumental Phase Errors
}

In VLBI observations, instrumental phase errors are caused 
by changes in the electrical path lengths in transmitting 
cables, gravity deformation of antenna structures, depending on the 
observing elevation, and independent frequency standards controlling 
VLBI station clocks. Those phase errors usually show slow systematic 
drifts, depending on the ambient temperature and the observing 
elevation angles; it is difficult to predict this behavior from 
physical models with the accuracy required for the VLBI data 
analysis. Phase referencing observations with fast antenna switching 
for a closely located pair of sources can cancel almost all of the 
instrumental phase errors. In this report we do not consider 
the contributions of such phase errors. 

\subsection{
  Uncertainties in the Positions and Structures of Calibrators
}

If there is little information about the position and structure 
of the chosen phase referencing calibrator, fringe phase errors 
are induced in the phase compensation. The residual phase error, 
$\sigma_{\phi_{\mathrm{\Delta s}}}$, due to the positional error 
of a calibrator, $\Delta s^{\mathrm{c}}$, is approximated as 
\begin{eqnarray}
\label{equ:03-19}
\sigma_{\phi_{\mathrm{\Delta s}}}~\mathrm{[rad]}
       &\approx&
    \frac{\sqrt{2}\pi\nu\Delta\theta B{\Delta}s^{\mathrm{c}}}{c}. 
\end{eqnarray}
The positions of celestial objects are determined in the 
International Celestial Reference Frame (ICRF; 
\cite{Ma1998}). 
The ICRF is defined by a set of extragalactic radio sources 
(ICRF sources) with simple and/or well-known structures  
(\cite{Ma1998}; 
\cite{Fey2004}). 
The ICRF sources have a typical astrometric accuracy of 0.3~mas 
at $S$-band (2.3~GHz) and $X$-band (8.4~GHz) 
based on the aggregation of geodetic and astrometric VLBI 
observations. Although the ICRF sources are well distributed 
over the sky, the sky coverage is not sufficient to provide 
suitable phase referencing calibrators. To make up for this, 
astrometric VLBI survey activities have progressed to find 
new phase referencing calibrators 
\citep{Beasley2002}.  
A typical astrometric accuracy of the new calibrator candidates 
is better than 5~mas 
\citep{Petrov2006}. 
The astrometric accuracy of those sources will be improved by 
repeated astrometric observations in the future. 
Equation~(\ref{equ:03-19}) can be expressed with the following 
approximation: 
\begin{eqnarray}
\label{equ:03-20}
\sigma_{\phi_{\mathrm{\Delta s}}}~{\mathrm{[deg]}}
      &\approx&
          46 \cdot 
          \left(
            \frac{\nu~\mathrm{[GHz]}}{43~\mathrm{GHz}}
          \right)
          \left(
            \frac{\Delta\theta~\mathrm{[deg]}}{2^\circ}
          \right)
                     \nonumber \\
      &&\times
          \left(
               \frac{B~\mathrm{[km]}}{25,000~\mathrm{km}}
          \right)
          \left(
               \frac{\Delta s^{\mathrm{c}}~\mathrm{[mas]}}{0.3~\mathrm{mas}} 
          \right).
\end{eqnarray}
The residual phase errors with the typical parameters used 
in equation~(\ref{equ:03-20}) are given in table~\ref{tbl:03-01}. 

There may be very few unresolved sources at space VLBI resolution. 
There is another possibility to use galactic water maser or 
silicon monoxide maser sources as phase referencing calibrators, 
but these galactic maser emissions may also be seriously resolved 
at VSOP-2 spatial resolution 
\citep{Migenes1999}. 
Because the contributions of source structures to the phase errors 
are dealt with case by case, we do not discuss them in this report. 

\subsection{
  Thermal Noise
}

The phase error, $\sigma_{\Delta\epsilon_{\mathrm{therm}}}$, 
due to thermal noise in phase referencing for a pair of VLBI 
telescopes is given by 
\begin{eqnarray}
\label{equ:03-17}
\sigma_{\Delta\epsilon_{\mathrm{therm}}}~[\mathrm{rad}]
                &=& \frac{\sqrt{2}\ k\ \overline{T_{\mathrm{sys}}}}
                         {\eta \overline{A_{\mathrm{e}}}
                          \sqrt{\Delta \nu}}
                      \sqrt{
                          \frac{1}{S^{\mathrm{t}2} T^{\mathrm{t}}}
                        + \frac{1}{2S^{\mathrm{c}2} T^{\mathrm{c}}}
                      }\ \ \ \ ,
\end{eqnarray}
where 
$k$ is Boltzmann's constant, 
$T_{\mathrm{sys}}$ is the system noise temperature, 
$A_{\mathrm{e}}$ is the effective aperture, 
$\eta$ is the coherence factor for the VLBI digital data processing, 
and $\Delta\nu$ is the observing IF bandwidth; 
$S^{\mathrm{t}}$ and $S^{\mathrm{c}}$ are the flux densities 
of the target and calibrator, respectively; 
$T^{\mathrm{t}}$ and $T^{\mathrm{c}}$ are the scan durations 
of the target and calibrator, respectively. 
Bars over the parameters represent the geometric mean 
of two telescopes. 
The factor of 
1/2 
of $1/(S^{\mathrm{c}2} T^{\mathrm{c}})$ comes 
from the interpolation process in which two neighboring calibrator 
scans are used for the phase compensation of a target scan 
between the two. Assuming two-bit sampling in analogue-to-digital (A/D) 
conversions and using a system equivalent flux density (SEFD), 
represented by $2kT_{\mathrm{sys}}/A_{\mathrm{e}}$, 
we obtain the following approximation: 
\begin{eqnarray}
\label{equ:03-18}
\sigma_{\Delta\epsilon_{\mathrm{therm}}}&&~\mathrm{[deg]}
          =   1.6\times10^{-5} \cdot
                \left(
                      \frac{\Delta\nu~\mathrm{[MHz]}}
                                 {256~\mathrm{MHz}}
                \right)^{-\frac{1}{2}}
                                   \nonumber \\
         &&\times
                \left[
                  \left(
                    \frac{T^{\mathrm{t}}~\mathrm{[s]}}{10~\mathrm{s}}
                  \right)^{-1}
                  \left(
                    \frac{\overline{S_{\mathrm{SEFD}}}~\mathrm{[Jy]}}
                         {S^{\mathrm{t}}~\mathrm{[Jy]}}
                  \right)^2
                \right .
        \nonumber \\
             &&+
                \left .
                 \left(
                    \frac{2T^{\mathrm{c}}~\mathrm{[s]}}{10~\mathrm{s}}
                  \right)^{-1}
                  \left(
                    \frac{\overline{S_{\mathrm{SEFD}}}~\mathrm{[Jy]}}
                         {S^{\mathrm{c}}~\mathrm{[Jy]}}
                  \right)^2
                  \right]^{\frac{1}{2}},
\end{eqnarray}
where 
$\overline{S_{\mathrm{SEFD}}}$ is the geometric mean of the 
SEFDs of two telescopes. 

\begin{table}
\caption{
Residual phase errors of a space baseline for a $2^\circ$ 
separate pair of sources. 
}\label{tbl:03-01}
\begin{center}
\begin{tabular}{p{17mm}p{16mm}p{16mm}p{16mm}}
\hline
\hline
Error item
   & \multicolumn{3}{c}{RMS phase} \\
\cline{2-4}

   & 8.4~GHz
   & 22~GHz
   & 43~GHz     \\
\hline
    $\sigma_{\phi_{\mathrm{dtrp}}}$\footnotemark[$*$,$\dagger$,$\ddagger$]
  & $12^\circ$
  & $31^\circ$
  & $61^\circ$ \\
    $\sigma_{\phi_{\mathrm{dion}}}$\footnotemark[$*$,$\dagger$,$\S$]
  & $ 4^\circ$
  & $ 1^\circ$
  & $ 1^\circ$ \\
    $\sigma_{\phi_{\mathrm{strp}}}$\footnotemark[$\dagger$,$\P$]
  & $15^\circ$
  & $40^\circ$
  & $78^\circ$ \\
    $\sigma_{\phi_{\mathrm{sion}}}$\footnotemark[$\dagger$,$\Vert$]
  & $25^\circ$
  & $ 9^\circ$
  & $ 5^\circ$ \\
    $\sigma_{\phi_{\mathrm{bl}}}$\footnotemark[$**$,$\dagger\dagger$]
  & $10^\circ$
  & $27^\circ$
  & $52^\circ$ \\
    $\sigma_{\phi_{\mathrm{\Delta s}}}$\footnotemark[$**$,$\ddagger\ddagger$]
  & $ 9^\circ$
  & $24^\circ$
  & $46^\circ$ \\
\hline
  \multicolumn{3}{@{}l@{}}{\hbox to 0pt{\parbox{85mm}{\footnotesize
      \par\noindent
      \footnotemark[$*$]
        Switching cycle time of 60~s. 
      \par\noindent
      \footnotemark[$\dagger$]
        Elevation angle of $45^\circ$ at the ground surface. 
      \par\noindent
      \footnotemark[$\ddagger$]
        Typical tropospheric condition. 
        ($C_{\mathrm{w}}=2$)
      \par\noindent
      \footnotemark[$\S$]
        50-percentile ionospheric condition. 
        ($C_{\mathrm{i}}=1.6\times10^{-5}$~TECU$\cdot$m$^{-5/6}$)
      \par\noindent
      \footnotemark[$\P$]
        Tropospheric zenith excess path error of 3~cm. 
      \par\noindent
      \footnotemark[$\Vert$]
        VTEC error of 6~TECU. 
      \par\noindent
      \footnotemark[$**$]
        Baseline length of 25000~km. 
      \par\noindent
      \footnotemark[$\dagger\dagger$]
        Antenna position error of 1~cm for a TRT, 
        OD error of 3~cm for an SRT, and 
        EOP error of 0.2~mas. 
      \par\noindent
      \footnotemark[$\ddagger\ddagger$]
        Calibrator positional error of 0.3~mas. 
    }\hss}}
\end{tabular}
\end{center}
\end{table}
\section{
  Simulation Study for VSOP-2 Phase Referencing
}

There are some conflicting issues in phase referencing. 
For example, the geometrical and atmospheric systematic 
errors causing an image distortion can be reduced by selecting 
a closer calibrator. On the other hand, brighter calibrators are 
often preferable because the larger is the thermal noise in the 
calibrator fringe, the less successful will the phase connection be. 
If there are several calibrator candidates around a target, 
which is better for phase referencing, the closer, but fainter, 
or the brighter, but more distant? In the end, the important 
thing is to select the optimum combination at the observing 
frequency in order to make the residual phase errors as small 
as possible. Constraints on the separation angle or the switching 
cycle time for a single space baseline can be evaluated from 
the approximations described in the previous section. However, 
for the image synthesis with a large amount of ($u$,~$v$) samples 
with multi-baselines, it is hard to predict the image quality from 
the approximations. In addition, cycle skips in the phase connection 
of the calibrator fringe phases, often occur as the observing frequency 
becomes higher, and/or as the switching cycle time 
becomes longer. Degradation in the image quality due to 
the cycle skips can hardly be predicted by an analytical method. 

In order to verify the effectiveness of phase referencing with VSOP-2, 
we developed a simulation tool called ARIS (Astronomical 
Radio Interferometor Simulator). In this section 
we first introduce what and how VLBI errors are simulated. 
Demonstrations of VSOP-2 phase referencing observations 
performed with ARIS are also shown. We then focus on the 
imaging performance in VSOP-2 phase referencing under 
realistic conditions so as to determine the allowable 
observing parameters, such as the switching cycle time, 
the separation angle, the OD accuracy of the satellite, 
the tropospheric condition, and the calibrator flux density. 

\subsection{
  Simulations of VLBI Fringe Errors with ARIS
}

\subsubsection{
  Atmospheric fringe phase errors
  \label{TEC_DATA_ANALYSES}
}

\begin{figure}
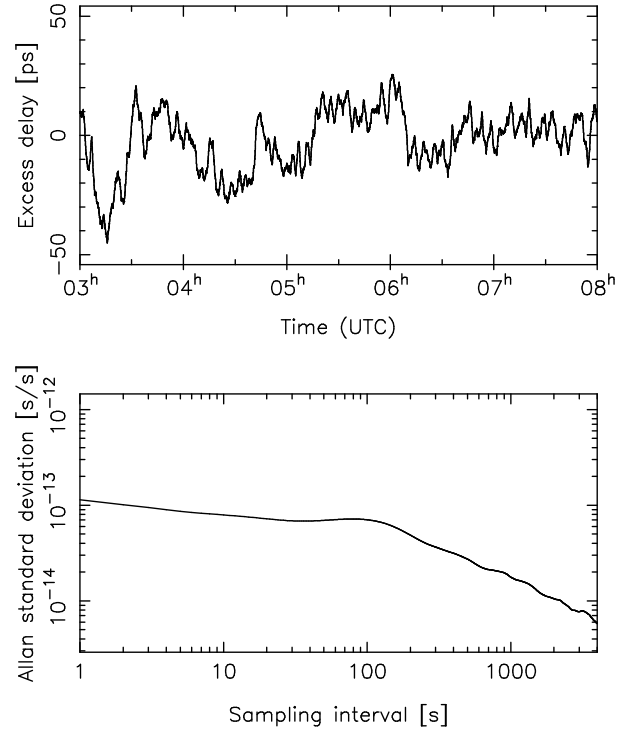

\begin{center}
    \FigureFile(80mm,80mm){figure-02.epsi}
\caption{
\label{fig:04-01}
Top: Typical simulated time series of the tropospheric excess path delay 
fluctuations of a space baseline under typical tropospheric 
conditions. The abscissa 
is time, and the ordinate is the excess path delay. 
The expectation is adjusted to 0 when the time 
series is generated. 
Bottom: Allan standard deviation calculated 
from the simulated time series shown at the top. 
The abscissa is the interval time, and the ordinate is 
the Allan standard deviation. 
}
\end{center}
\end{figure}

In ARIS the tropospheric phase fluctuations are modeled as 
a phase screen assuming Kolmogorov turbulence. This simple 
model is useful when considering the interferometric 
phase fluctuations due to the troposphere 
\citep{Asaki1996}. 
The grid interval of the screen is set to 1~m. Since, in ARIS, 
the inner and outer scales can be selected among 
$2^p$~$\times$~(grid interval), where $p$ is a natural number, 
those scales are fixed to 1024 and 8192~m for the inner and 
outer scales, respectively. 
The phase screen is simulated for each terrestrial telescope site, 
and flows at a constant speed of 10~m~s$^{-1}$ along west wind. 
When one screen passes over the line-of-sight of a terrestrial 
telescope, another new one is created from the edge of the previous 
screen as a seed so as not to generate an unnatural gap between them. 
The altitude of the phase screen is 1~km, and the elevation dependence 
of the amplitude of the fluctuation is achieved by multiplying 
a factor of $\sqrt{\sec{Z_{\mathrm{g}}}}$ as shown 
in equation (\ref{equ:03-05}). 
A typical simulated time series of the tropospheric fluctuations and 
the Allan standard deviation are shown in figure~\ref{fig:04-01}. 

For ionospheric phase fluctuations, we consider two components 
in the TEC fluctuations: 
one is the MS-TID, and the other is a phase screen and assuming 
Kolmogorov turbulence driven by the MS-TID. The former is 
sinusoidal waves with a spatial wave length of 200~km and 
a propagating speed of 100~m~s$^{-1}$ at an altitude of 300~km. 
The latter has a grid interval of 150~m, and flows at the same 
speed and at the same altitude of the MS-TID. The inner scale 
of the screen is set to 76.8~km so as not to be over the spatial 
wave length of the MS-TID, and no transition region from the inner 
to outer scales of the SSF is provided in ARIS. Although there 
can be cases in which a single ionospheric phase screen covers 
separate terrestrial telescopes, the phase screen is independently 
simulated for each terrestrial telescope site. 

\begin{figure}
\begin{center}
\begin{tabular}{c}
    \FigureFile(80mm,80mm){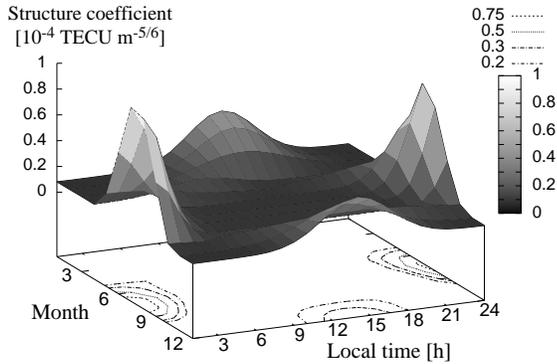}
\end{tabular}
\end{center}
\caption 
{\label{fig:04-02}
Amplitude of the structure coefficient of the SSF of 
the TEC fluctuations versus local time and season for 
the northern hemisphere to be adopted in ARIS. 
}
\end{figure}

\begin{figure}
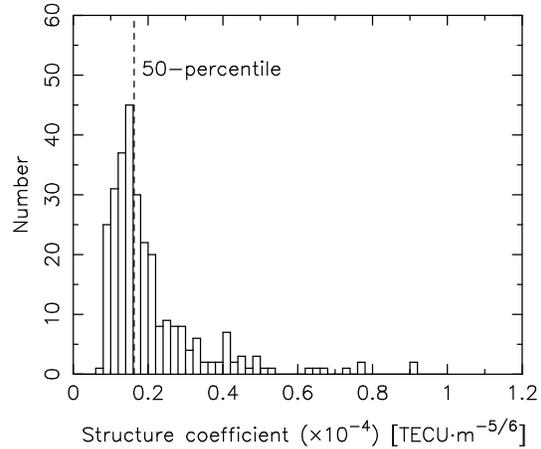

\begin{center}
\begin{tabular}{c}
    \FigureFile(70mm,70mm){figure-04.epsi}
\end{tabular}
\end{center}
\caption 
{\label{fig:04-03}
Histogram of the structure coefficient of the SSF of the 
TEC fluctuations obtained from the GPS TEC data analysis. 
The data taken from 2001 January to December were analyzed 
with a method based on 
\citet{Noguchi2001}. 
The vertical dotted line shows the 50-percentile condition. 
}
\end{figure}

\begin{figure}
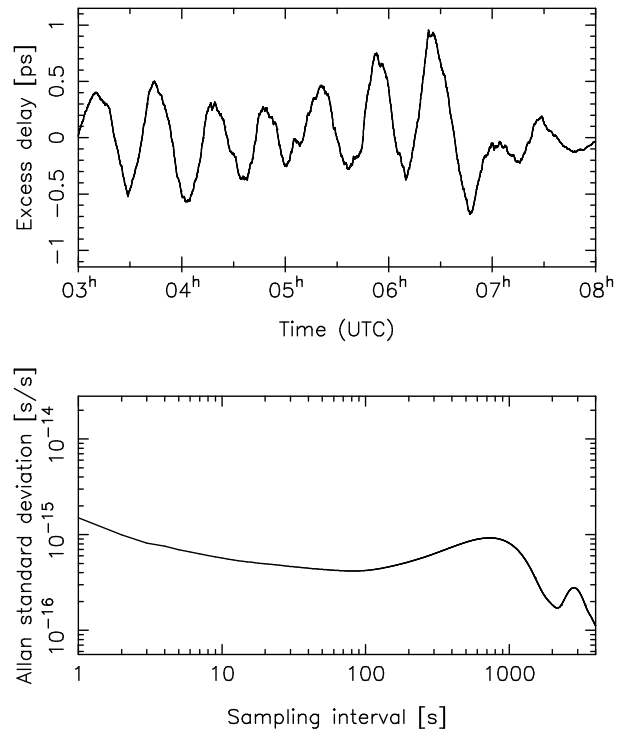

\begin{center}
\begin{tabular}{c}
    \FigureFile(80mm,80mm){figure-05.epsi}
\end{tabular}
\caption{
\label{fig:04-04}
Top: Typical simulated time series of the ionospheric excess path 
delay fluctuations of a space baseline in winter night time. 
The abscissa is time, and the ordinate 
is the excess path delay at 43~GHz. 
The expectation is adjusted to 0 when the time 
series is generated. 
Bottom: Allan standard deviation calculated 
from the simulated time series shown in the top. 
The abscissa is the interval time, and the ordinate is 
the Allan standard deviation in s/s. 
}
\end{center}
\end{figure}

The simulated MS-TID and phase screens are transmitted from 
the geographical poles to the equator along the longitudes. 
Recall that Kolmogorov turbulence is assumed to be driven 
by the MS-TID, so that the amplitude of Kolmogorov turbulence 
is proportional to that of the MS-TID. To determine the balance 
of the amplitudes between them, we analyzed short-term 
VTEC fluctuations, based on 
\citet{Noguchi2001}, 
of TEC data taken with a Japanese GPS receiver network operated 
by Geographical Survey Institute (GSI) of Japan 
\citep{Saito1998}. 
In this study we kept the balance so that $C_{\mathrm{i}}$ has 
its maximum value of $10^{-4}$~TECU$\cdot$m$^{-5/6}$ when 
the amplitude of the MS-TID is 1~TECU, the largest MS-TID amplitude 
often observed at mid-latitudes. Since the influence of 
the ionospheric phase fluctuations is generally smaller 
than that of the tropospheric phase fluctuations at the 
VSOP-2 observing frequency bands, we did not pay attention 
to various ionospheric conditions. Instead, the time evolution 
of the structure coefficient versus the local time and season 
for the northern hemisphere, as shown in figure~\ref{fig:04-02}, 
is given. This figure was obtained from an analysis based on 
\citet{Noguchi2001} 
with a smoothing process by means of an elliptical Gaussian fitting. 
From the GPS TEC analysis, the 50-percentile condition of 
$C_{\mathrm{i}}$ is 1.6$\times10^{-5}$~TECU~m$^{-5/6}$, 
as shown in figure~\ref{fig:04-03}. 
In ARIS, an unusual ionospheric status, such as a large-scale TID, 
plasma bubbles often observed at low-latitudes, and geomagnetic storms, 
is not considered. The elevation dependence of the amplitude of the 
fluctuations is achieved by multiplying a factor of 
$\sqrt{\sec{Z_{\mathrm{i}}}}$. 
The simulated time series of the ionospheric fluctuations at 43~GHz, 
observed with a space baseline, and the Allan standard deviation are 
shown in figure~\ref{fig:04-04}. 


The atmospheric systematic error was set randomly and independently 
for each terrestrial telescope in each simulation pass. 
The standard deviations were 3~cm and 6~TECU for the systematic errors 
of the zenith excess path length due to the water vapor and VTEC, 
respectively. The line-of-sight tropospheric excess path error 
is a product of the zenith path error and Niell's wet mapping function 
\citep{Niell1996}. 
The line-of-sight TEC error is a product of the VTEC error and 
$\sec{Z_{\mathrm{F}}}$. The altitude to calculate $Z_{\mathrm{F}}$ 
was set to 450~km. 

\subsubsection{
  Geometrical phase errors and satellite trajectory in ($u$,~$v$)
  \label{OD_MODEL}
}

In ARIS a displacement in the a priori terrestrial telescope position 
adopted in the correlator is given as a displacement vector. 
The standard deviations of the horizontal and vertical components 
are 3~mm and 1~cm, respectively, in the local tangent coordinates, 
and the displacement vector is set randomly and independently 
for each terrestrial telescope in each simulation pass. 
The EOP errors are given as offsets in the EOP transformation matrices 
defined by IERS Conventions (2003) 
(\cite{McCarthy2004}\footnote{
IERS Conventions, No.32, ch5 (2004) is available at 
$<$http://www.iers.org/iers/publications/tn/tn32/$>$.}). 
The offset of the celestial and terrestrial polar directions are 
randomly set with standard deviations of 0.3~mas, and the 
UT1 offset is randomly set with standard deviation of 0.02~ms. 
The systematic error of calibrator positions is given with the 
standard deviation of 0.3~mas, while the positional offset is 
logged in ARIS and used to estimate the astrometric accuracy 
of the target in the later analysis stage. 

\begin{figure}
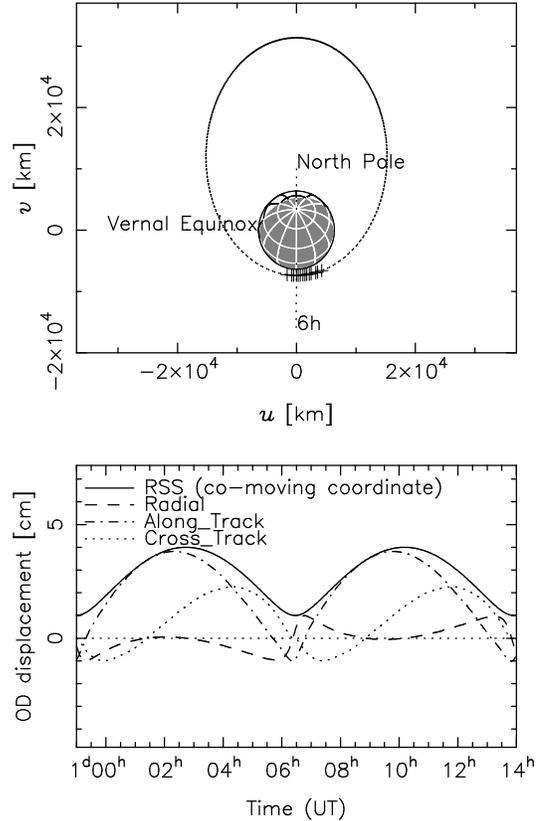

\begin{center}
\begin{tabular}{c}
    \FigureFile(70mm,70mm){figure-06.epsi}
\end{tabular}
\end{center}
\caption 
{\label{fig:04-05}
Top: Simulated orbital trajectory of the VSOP-2 satellite 
projected to the ($u$,~$v$) plane at 
($6^{\mathrm{h}}$,~$59^\circ$). 
Dots and crosses represent the trajectory out of and in the Earth 
shadow, respectively. The Earth size is represented by the central 
circle with the gray part of night.
Bottom: Simulated OD displacement from the supposed 
trajectory of the satellite in the co-moving coordinates. 
}
\end{figure}

In ARIS the displacement vector of the satellite position is 
modeled in a co-moving frame: two unit vectors are aligned with 
the radial direction towards geocenter (radial component) 
and the orbital angular momentum vector (cross track component); 
the remained component is in the orbital plane 
(along track component) and orthogonal to the other two. 
The three components of the displacement vector 
are made from a combination of ad~hoc trigonometric functions 
synchronizing with the true anomaly of the orbit. 
Since the OD accuracy would be the worst at apogee, 
as described in subsection \ref{OD_METHOD}, 
the root-square-sum (RSS) of the three components of the displacement 
at apogee is fixed to be four-times larger than that at perigee. 
The displacement vector is scaled up or down in order to investigate 
the influence of the OD error on the image quality. In the following 
discussion we use the OD displacement at the apogee (ODDA) 
as an indicator of the OD error. 
Figure~\ref{fig:04-05} shows a trajectory of the satellite 
orbital motion and the OD displacement from the supposed 
trajectory in the case of the ODDA of 4~cm. 

\subsubsection{
  How to generate corrupted fringes
}

The instantaneous complex visibility, 
$\mbox{\boldmath $V$}(i, j, t, \nu_0)$, for a baseline of 
telescopes $i$ and $j$ at time $t$ and at the sky frequency 
at the IF band edge, $\nu_0$, is calculated from 
a two-dimensional Fourier transformation of a brightness distribution 
of a celestial source and the ($u$,~$v$) trajectory every second. 
We assume that the delay difference and fast phase change 
due to the Earth rotation are largely removed in the correlator. 
The multi-channel frequency, $\nu$, in the IF bandwidth is 
$\nu_0 + (m/M)\Delta\nu$, 
where $M$ is the number of the frequency channels, 
and $m$ is 0, 1, 2, ..., $M-1$. 
In ARIS the non-dispersive delay errors listed in 
table~\ref{tbl:03-01} (all but the ionospheric excess delay) 
are calculated for each telescope. Those delay errors are 
summed to obtain a total non-dispersive delay error, 
$\tau_{\mathrm{err1}}(i, t)$. 
On the other hand the ionospheric delay error, 
$\tau_{\mathrm{err2}}(i, t, \nu)$, 
for a terrestrial telescope is calculated with the 
line-of-sight TEC estimation error $\Delta I$ by 
\begin{eqnarray}
\label{equ:04-01}
\tau_{err2}(i, t, \nu) &=& -\frac{\kappa TEC_{\mathrm{err}}}{c\nu^2}. 
\end{eqnarray}
Then, the contribution of the delay errors to the antenna 
complex gain, $\mbox{\boldmath $G$}(i, t, \nu)$, 
is obtained by 
\begin{eqnarray}
\label{equ:04-02}
\mbox{\boldmath $G$}(i, t, \nu) &=&
  \exp{
    \{
        -j2\pi\nu[
            \tau_{\mathrm{err1}}(i, t)
          + \tau_{\mathrm{err2}}(i, t, \nu)
        ]
    \}
  }.
\end{eqnarray}
Thus, a fringe $\mbox{\boldmath $F$}(i, j, t, \nu)$ of 
the source for the baseline is made by 
\begin{eqnarray}
\label{equ:04-03}
  \mbox{\boldmath $F$}    (i, j, t, \nu) &=& 
  \mbox{\boldmath $G$}    (i,    t, \nu)
  \mbox{\boldmath $G$}^{*}(   j, t, \nu)
  \mbox{\boldmath $V$}    (i, j, t, \nu_0),
\end{eqnarray}
where the asterisk represents complex conjugate, and 
$\mbox{\boldmath $V$}$ is assumed not to vary with 
the frequency channels. A Gaussian noise vector, 
whose components correspond to the real and imaginary parts, is 
generated every second at each frequency channel for each baseline. 
Both components 
have an RMS amplitude of  
$\overline{S_{\mathrm{SEFD}}}(i, j, t, \nu_0)/
{\mid\mbox{\boldmath $V$}(i, j, t, \nu_0)\mid}/
(\eta\sqrt{2\Delta\nu/M})$, 
where 
$\overline{S_{\mathrm{SEFD}}}(i, j, t, \nu_0)$ 
is the geometric mean of the SEFDs of the telescopes, $i$ and $j$ 
at the observing elevations at $\nu_0$, and assumed not to be 
varied with the frequency channels. 
The fringe 
including the thermal noise is generated by 
adding the noise vector. 
The parameters to simulate $\tau_{\mathrm{err1}}$ and 
$\tau_{\mathrm{err2}}$ are listed in table~\ref{tbl:04-01}. 
Since the phase screen generation is the most time-consuming process 
in ARIS, the time-variation patterns of the atmospheric phase 
fluctuations are repeatedly used for various observing frequencies, 
ODDsA, and $C_{\mathrm{w}}$. 

\begin{table}
\caption{
Parameters used in the simulations. 
}\label{tbl:04-01}
\begin{center}
\begin{tabular}{lcl}
\hline
\hline
Item
  & Type\footnotemark[$*$]
  & Quantity   \\
\hline
$\Delta\theta$
  & V
  & 0.5, 1, 2, 4, 8$^\circ$ \\
ODDA
  & V
  & 2, 4, 8, 16, 32~cm \\
$\nu$
  & V
  & 8.4, 22, 43~GHz \\
$C_{\mathrm{w}}$      
  & V
  & 1, 2, 4 \\
\hline
$v_{\mathrm{w}}$ (west wind)
  & C
  & 10~m~s$^{-1}$   \\
$H_{\mathrm{w}}$   
  & C
  & 1~km    \\
$H_{\mathrm{s}}$   
  & C
  & 1~km    \\
\hline
$C_{\mathrm{i}}$   
  & $\cdots$\footnotemark[$\dagger$]
  & $\cdots$\footnotemark[$\dagger$] \\
$v_{\mathrm{i}}$ 
  & C
  & 100~m~s$^{-1}$    \\
$H_{\mathrm{i}}$ 
  & C
  & 300~km    \\
\hline
$\Delta l_{\mathrm{z}}$
  & T
  & 3~cm     \\
\hline
$\Delta VTEC$  
  & T
  & 6~TECU   \\
$H_{\mathrm{F}}$ 
  & C
  & 450~km    \\
\hline
$\Delta P_{\mathrm{TRT}}$ (vertical) 
  & T
  & 1~cm   \\
\ \ \ \ \ \ \ \ \ \ (horizontal)
  & T
  & 3~mm   \\
\hline
Celestial pole offset
  & S
  & 0.3~mas \\
Terrestrial pole offset
  & S
  & 0.3~mas \\
UT1 offset
  & S
  & 0.02~ms \\
\hline
$\Delta s^{\mathrm{c}}$
  & S
  & 0.3~mas \\
\hline
  \multicolumn{3}{@{}l@{}}{\hbox to 0pt{\parbox{85mm}{\footnotesize
      \par\noindent
      \footnotemark[$*$]
        Type~V items are interesting ones to verify the effectiveness 
        of VSOP-2 phase referencing. 
        Type~C items have a constant value. 
        Type~T items are set for each telescope in each simulation 
        pass among samples with the mean of 0 and the standard 
        deviation given by ``Quantity". 
        Type~S items are set in each simulation pass among samples 
        with the mean of 0 and the standard deviation given 
        by ``Quantity". 
      \par\noindent
      \footnotemark[$\dagger$]
        See subsubsection \ref{TEC_DATA_ANALYSES}.
    }\hss}}
\end{tabular}
\end{center}
\end{table}

\subsubsection{
  How to compensate for corrupted fringe phases
}

\begin{figure}
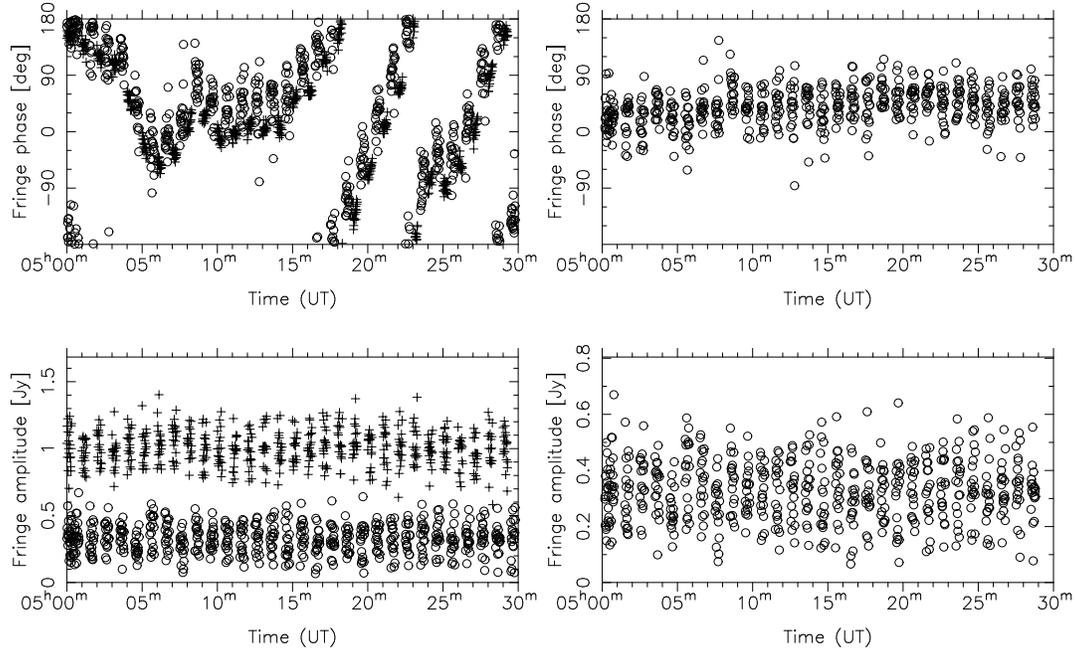

\begin{center}
\begin{tabular}{c}
    \FigureFile(70mm,70mm){figure-07a.epsi}
    \FigureFile(70mm,70mm){figure-07b.epsi}
\end{tabular}
\end{center}
\caption 
{\label{fig:04-06}
Simulated time series of the fringes of a VSOP-2 space baseline 
with a VLBA telescope at 43~GHz under good tropospheric 
conditions. The IF bandwidth is 256~MHz, 
the OD displacement at the apogee is 4~cm, the switching cycle time is 
60~seconds, and the separation angle of the sources is $2^{\circ}$.  
The target and calibrator flux densities are 0.3 and 1~Jy, 
respectively. 
Left: Fringe phases (top) and amplitudes (bottom) 
of the target and calibrator. Open circles and crosses 
represent the target and calibrator fringes, respectively. 
Right: Fringe phase (top) and amplitude (bottom) 
of the target after phase referencing. 
}
\end{figure}

The phase-compensation process is performed in ARIS 
with a simple interpolation algorithm. 
The phase-compensated fringe phases are first vector-averaged for 
each scan. The calibration data, $\Phi_{\mathrm{cal}}$, 
for the $n$-th target scan is 
\begin{eqnarray}
\label{equ:04-05}
\Phi_{\mathrm{cal}}(t) &=& \overline{\Phi^{\mathrm{c}}_{n}} + 
    \frac{
      \left(
        \overline{\Phi^{\mathrm{c}}_{n+1}}
      - \overline{\Phi^{\mathrm{c}}_{n}} + \Phi_{\mathrm{adj}} 
      \right)
      (t - t^{\mathrm{c}}_{n}) 
    }{T_{\mathrm{swt}}},
\end{eqnarray}
where $\overline{\Phi^{\mathrm{c}}_{n}}$ is a vector-averaged 
calibrator fringe phase along the time and frequency followed 
by the $n$-th target scan, 
$\overline{\Phi^{\mathrm{c}}_{n+1}}$ is the following one, and 
$t^{\mathrm{c}}_{n}$ is the center time of the followed 
calibrator scan. An adjustment term, $\Phi_{\mathrm{adj}}$, 
is selected among $-2\pi$, 0, and $+2\pi$ radians to prevent 
$\mid\overline{\Phi^{\mathrm{c}}_{n+1}}
   - \overline{\Phi^{\mathrm{c}}_{n}}\mid$ 
from being over $\pi$ radians, so as not to induce a cycle skip 
between the neighboring calibrator scans. The compensated target 
fringe, 
$\mbox{\boldmath $F$}^{\prime\mathrm{t}}(i, j, t, \nu)$, is 
calculated from 
$\mbox{\boldmath $F$}^{\mathrm{t}}(i, j, t, \nu)$ and 
$\Phi_{\mathrm{cal}}(t)$ by 
\begin{eqnarray}
\label{equ:04-06}
  \mbox{\boldmath $F$}^{\prime\mathrm{t}}(i, j, t, \nu) &=& 
  \mbox{\boldmath $F$}^{\mathrm{t}}(i, j, t, \nu)
                           \exp{[-j\Phi_{\mathrm{cal}}(t)]}. 
\end{eqnarray}
Figure~\ref{fig:04-06} shows a demonstration of a phase 
referencing observation at 43~GHz, performed with ARIS, 
of a baseline consisting of the VSOP-2 satellite and 
a terrestrial telescope of the Very Long Baseline Array (VLBA) 
of National Radio Astronomy Observatory (NRAO) 
under good tropospheric conditions ($C_{\mathrm{w}}=1$). 
The switching cycle time and scan durations for both the 
target and calibrator are 60 and 20~seconds, respectively. 
Both are point sources, and the separation angle is $2^\circ$. 
The target and calibrator flux densities are 0.3 and 1~Jy, 
respectively. The ODDA is 4~cm. In this demonstration 
the coherence of the target fringe phase is improved 
by phase referencing from a situation including the 
prominent perturbations to an RMS phase of $30^\circ$. 

\onecolumn
\twocolumn

\subsubsection{
  Observational limits on the satellite
}

\begin{figure}
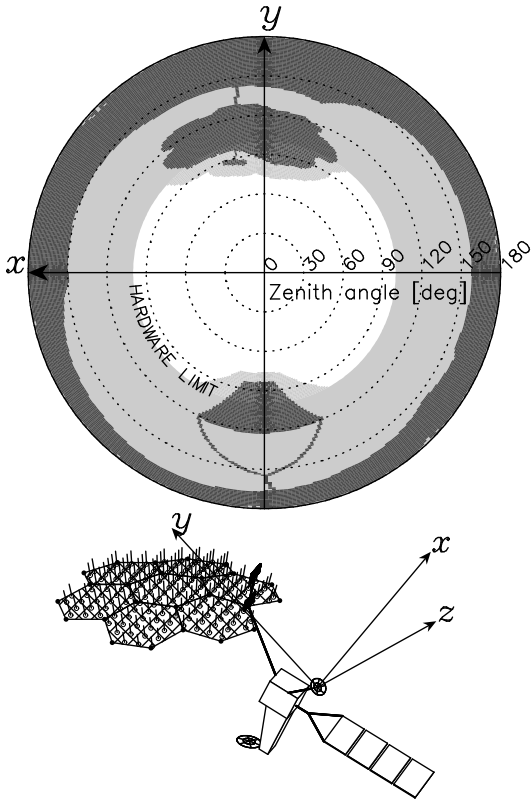

\begin{center}
\begin{tabular}{c}
    \FigureFile(70mm,70mm){figure-08.epsi}
\end{tabular}
\end{center}
\caption 
{\label{fig:04-07}
Field of view of the data-link antenna based on the 
conceptual design of the satellite. 
The elevation limit of the data-link antenna is 
$-10^\circ$ for the time being in this investigation. 
A light-gray region represents the masked area, because 
of the assumed elevation limit of the data-link antenna 
and satellite structure (dark gray region) in the near field.
}
\end{figure}

The VSOP-2 observing time for phase referencing will be 
restricted by power-supply shortages of the satellite 
because of Earth eclipse, and inaccessible data-link 
because of the limited distribution of the ground tracking network 
and angular coverage of the on-board data-link antenna. 
In order to realistically simulate the ($u$,~$v$) coverage 
with the space baselines, the following three conditions are 
considered: 
(1) phase referencing observations are suspended during partial 
and total Earth eclipses in the satellite orbit; 
(2) the ground tracking network used in VSOP 
\citep{Hirabayashi2000} 
is assumed for VSOP-2; 
and 
(3) on-board link antenna coverage is tentatively generated 
from a blocking pattern based on the conceptual design 
of the satellite, as shown in figure~\ref{fig:04-07}. 

\subsection{
  Allowable Separation Angle, Switching Cycle Time, 
  and OD Error for VSOP-2 Phase Referencing
}

Because fast attitude switching of the VSOP-2 satellite for 
phase referencing will be achieved by actuator attitude control, 
we have to clarify the requirements of the control system 
in terms of the maneuvering speed and the angular range. 
The OD accuracy is also an issue to be discussed for the 
satellite system. Therefore, it is important to clarify the 
allowable separation angle, switching cycle time, and OD error 
so as to determine the satellite design. Since the allowable 
switching cycle time is very dependent on the tropospheric condition 
and observing frequency, we must investigate the performance 
of phase referencing under various tropospheric conditions 
at each observing frequency. Here, we present a simulation series 
of VSOP-2 phase referencing observations for various separation 
angles, ODDsA, switching cycle times, tropospheric conditions, 
and observing frequencies. 

\subsubsection{
  Simulation conditions
}
\begin{figure}
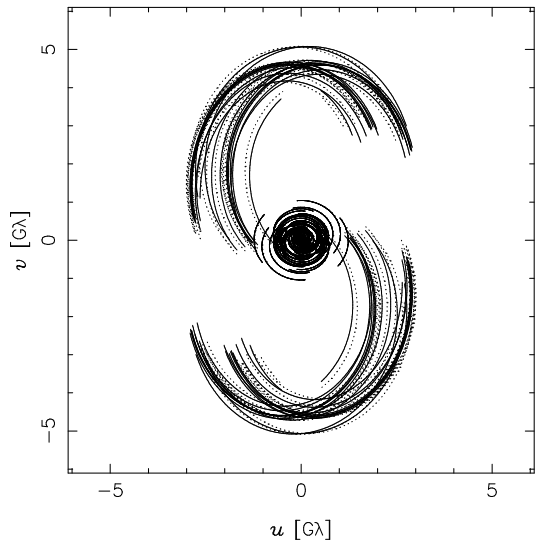

\begin{center}
\begin{tabular}{c}
    \FigureFile(70mm,70mm){figure-09.epsi}
\end{tabular}
\end{center}
\caption 
{\label{fig:04-08}
Example of ($u$,~$v$) coverage of a VSOP-2 observation at 43~GHz 
for a $2^\circ$ separate pair of sources along with Right Ascension. 
The terrestrial array is 10 VLBA telescopes. Solid and 
dashed lines represent the target and calibrator, respectively. 
}
\end{figure}

The simulation conditions are summarized in table~\ref{tbl:04-02}. 
We simulated 15-hour observations with the VSOP-2 satellite and 
10 VLBA telescopes. This observing time is almost double of the 
planned orbital period of the satellite. The lowest elevation 
angle for the terrestrial telescopes was set to $20^\circ$. 
The center position of pairs of sources was at 
($6^{\mathrm{h}}$,~$59^\circ$). Such observing conditions were 
selected because 
(1) 
the highest spatial resolution in VSOP-2 is achievable; 
(2) 
long-term phase variations due to the geometrical and 
atmospheric systematic errors on the synthesized image 
can be tested; 
and 
(3) 
it is easy to investigate the influence of the sensitivity 
of the space baselines on the synthesized images by using a 
homogeneous terrestrial VLBI array. 
The SEFDs of the VLBA telescopes were taken from 
\citet{Napier1995} 
as well as 
J.~S.~Ulvestad and J.~M.~Wrobel 
(2006, VLBA Observational Status Summary), 
\footnote{
$<$http://www.vlba.nrao.edu/astro/genuse/$>$.} 
and are listed in table~\ref{tbl:04-03}. 
Figure~\ref{fig:04-08} shows an example of the ($u$,~$v$) 
coverage of the simulations for the case of a $2^\circ$ 
separate pair of sources. 
Although the variable ODDA with 
an expectation and distribution 
seems to be more realistic, much larger data sets will be needed 
to show the statistically reliable results in the current 
imaging simulations with only two orbits. 
To reduce the number of simulations in order to save the time, 
we adopted non-variable ODDsA of 2, 4, 8, 16, and 32~cm. 

\begin{table}
\caption{
  Source positions, observing schedule, terrestrial telescopes, 
  and orbital parameters. 
}\label{tbl:04-02}
\begin{center}
\begin{tabular}{ll}
\hline
\hline
Center position of sources     & RA(2000)$=$\timeform{6h00m00s}  \\
                               & Dec(2000)$=$\timeform{59D00'00"} \\
Observation start time         & 23h UT on Jan 1st \\
Observing time                 & 15 hours \\
\hline
Terrestrial array              & VLBA (10 telescopes) \\
Elevation limit of TRTs        & 20$^\circ$ \\
\hline
IF bandwidth                   & 256~MHz \\
A/D quantization level         & Two-bit sampling \\
Output IF band                 & 1 \\
Output frequency channel       & 1 \\
\hline
Inclination                    & $31^\circ$ \\
RA of the ascending node       & $180^\circ$ \\
Argument of perigee            & $-90^\circ$ \\
Apogee height                  & 25000~km \\
Perigee height                 & 1000~km \\
Time of perigee passage        & Observation start time  \\
\hline
\end{tabular}
\end{center}
\end{table}
\begin{table}
\caption{
SEFDs of the VSOP-2 SRT and the VLBA telescopes. 
}\label{tbl:04-03}
\begin{center}
\begin{tabular}{lll}
\hline
\hline
Telescope
  & Frequency
  & SEFD          \\
\hline

  & 8.4~GHz
  & 4708~Jy       \\
SRT
  & 22~GHz
  & 2353~Jy       \\

  & 43~GHz
  & 3766~Jy       \\
\hline

  & 8.4~GHz
  & 307~Jy\footnotemark[$\dagger$] \\
VLBA\footnotemark[$*$]
  & 22~GHz
  & 888~Jy\footnotemark[$\dagger$] \\

  & 43~GHz
  & 1437~Jy\footnotemark[$\dagger$] \\
\hline
  \multicolumn{3}{@{}l@{}}{\hbox to 0pt{\parbox{85mm}{\footnotesize
      \par\noindent
      \footnotemark[$*$] 
        \citet{Napier1995}; 
        J.~S.~Ulvestad and J.~M.~Wrobel (2006).
      \footnote{
        $<$http://www.vlba.nrao.edu/astro/genuse/$>$.
      }
      \par\noindent
      \footnotemark[$\dagger$]
        Estimate to the zenith direction. 
    }\hss}}
\end{tabular}
\end{center}
\end{table}

In ARIS, three tropospheric conditions of $C_{\mathrm{w}}=$1, 2, 
and 4, that is, good, typical, and poor tropospheric conditions 
were examined. The time evolution of $C_{\mathrm{w}}$ during an 
observation is not considered, and an identical tropospheric 
condition is used for all of the terrestrial telescopes. 
To compare the results for various values of the interesting 
parameters, we did not change 
the observation season, or start time to 
switch the tropospheric condition, because the other 
conditions, such as the ($u$,~$v$) coverage, should be identical 
throughout the simulations. The amplitude of the TEC disturbance 
varies along with the model, as shown in figure~\ref{fig:04-02}. 

In the present simulations the target and calibrator were 
point sources with the flux densities of $10^{-5}$ and $10^{-4}$ 
of $\overline{S_{\mathrm{SEFD}}}$, respectively, 
of a space baseline with a VLBA telescope directed to the 
zenith, as listed in table~\ref{tbl:04-04}. The flux densities 
of the target were 12, 14, and 23~mJy at 8.4, 22, and 43~GHz, 
respectively. While the VSOP-2 receivers have two IF bands 
each with 128~MHz in the conceptual design, 
the simulated observation system has, for simplicity, 
a single frequency channel with a bandwidth of 256~MHz. Although 
a coherence loss of about 10\% for the 256-MHz bandwidth occurs 
if a group delay error of 1~ns exists, the amplitude degradation 
was not considered in the present study. 
An independent phase drift in each IF band, which may cause 
a problem in their combination, was also not considered.  

\begin{table}
\caption{
  Flux densities of the target and calibrator in the
  first simulation series.
}\label{tbl:04-04}
\begin{center}
\begin{tabular}{llll}
\hline
\hline
Item        & 8.4~GHz & 22~GHz & 43~GHz \\
\hline
\vspace{-3mm}
 & & & \\
$\overline{S_{\mathrm{SEFD}}}$\footnotemark[$*$]
  & 1202 Jy
  & 1445 Jy
  & 2326 Jy \\
$S^{\mathrm{t}}$ ($\overline{S_{\mathrm{SEFD}}}\times10^{-5}$)
  &  12 mJy
  &  14 mJy
  &  23 mJy \\
$S^{\mathrm{c}}$ ($\overline{S_{\mathrm{SEFD}}}\times10^{-4}$)
  & 120 mJy
  & 145 mJy
  & 233 mJy \\
\hline
  \multicolumn{3}{@{}l@{}}{\hbox to 0pt{\parbox{120mm}{\footnotesize
      \par\noindent
      \footnotemark[$*$]
        Geometric mean of SEFDs of the SRT and a VLBA antenna to the zenith.
    }\hss}}
\end{tabular}
\end{center}
\end{table}

We conducted two cases for switching cycle times of 60 and 40~seconds. 
In the simulations, the slew time of all the telescopes was fixed 
to 10 seconds when changing the pointing for all the separation angles, 
so that the scan durations were 20 and 10 seconds for switching cycle 
times of 60 and 40 seconds, respectively. 
The expected signal-to-noise ratios for a single calibrator scan of 20 and 
10 seconds are 8.6 and 6.0, respectively, for the space 
baselines with assumptions of a bandwidth of 256~MHz, 
two-bit sampling in the A/D conversion, and 
the terrestrial telescope directed to the zenith. 

Baseline-based phase referencing were carried out in ARIS. 
The phase-compensated target fringes were stored in the FITS-IDI 
format to load into the NRAO Astronomical Image Processing System 
(AIPS). Phase-compensated fringes were Fourier inverted 
and deconvolved (CLEANed) with AIPS IMAGR to obtain 
synthesized images. A typical beam size at 43~GHz with 
uniform weighting was 80 and 38~$\mu$as for the major and minor 
axes, respectively, with a position angle of $86^\circ$. 
At 8.4 and 22~GHz the beam size increased by factors of 
5.1 and 2.0, respectively. We set the image pixel size to 
19, 7, and 3.8~$\mu$as at 8.4, 22, and 43~GHz, respectively. 
The synthesized image size was fixed to $256\times256$ pixels 
centered at the phase tracking center. A CLEAN box centered 
on the imaged area has a $100\times100$-pixel size. Here, we 
call the CLEAN image with the phase-compensated data 
a phase referencing direct image. 

Since the calibrator position was 
set with the randomly generated positional offset, 
the brightness distribution of the phase referencing 
direct image was shifted from the phase tracking center 
almost as much as the positional offset of the calibrator 
logged by ARIS. This positional shift from the origin 
is used to estimate the relative position of the target to 
the calibrator. If the imaged area 
is not wide enough to follow the positional shift by chance, 
the brightness distribution may go outside of the area. 
In the current simulations, plenty of data sets were 
processed, so that the iteration process with step-by-step 
inspections should be excluded to save time. 
Therefore, the phase tracking center of the compensated 
target data was automatically corrected with AIPS UVFIX 
by the coordinate rotation with the rotation axis 
$
\mbox{\boldmath $e$} =
\left [
(\mbox{\boldmath $s$}^{\mathrm{c}} +
 \mbox{\boldmath $s$}^{\mathrm{t}})
 \times
(\mbox{\boldmath $s$}^{\mathrm{c}} -
 \mbox{\boldmath $s$}^{\prime\mathrm{c}})
\right ] /
\mid
(\mbox{\boldmath $s$}^{\mathrm{c}} +
 \mbox{\boldmath $s$}^{\mathrm{t}})
 \times
(\mbox{\boldmath $s$}^{\mathrm{c}} -
 \mbox{\boldmath $s$}^{\prime\mathrm{c}})
\mid
$ and rotated angle 
$
\Psi = 
2 \arcsin{
\left [
\mid
 \mbox{\boldmath $s$}^{\mathrm{c}} -
 \mbox{\boldmath $s$}^{\prime\mathrm{c}}
\mid
 / 2\sqrt{1-(\mbox{\boldmath $s$}^{\mathrm{c}}\cdot\mbox{\boldmath $e$})^{2}}
\right ]
}
$, where 
$\mbox{\boldmath $s$}^{\mathrm{t}}$ 
and 
$\mbox{\boldmath $s$}^{\mathrm{c}}$ 
are unit vectors pointing to the target and calibrator, 
respectively, and 
$\mbox{\boldmath $s$}^{\prime\mathrm{c}}$ 
represents the calibrator position adopted in the correlator. 

Independent simulations with the same values of $C_{\mathrm{w}}$, 
ODDA, and separation angle were carried out for 16 pairs of sources 
with 16 different position angles spaced at $22^{\circ}.5$ intervals. 
The surface brightness peak value and its positional offset 
from the corrected phase tracking center were obtained in each 
image. Those results were averaged over the 16 different 
positions for statistical reliability. 

We used the ratio of the averaged brightness peak value 
to the model as an indicator to evaluate the resultant 
image quality without any dependence on the thermal noise of 
the target and ($u$,~$v$) coverage. 
We refer to the ratio as ``image coherence.'' 
The averaged positional offset was used as an indicator of 
astrometric accuracy of the phase referencing direct images. 

\subsubsection{
  Results
}

\begin{figure}
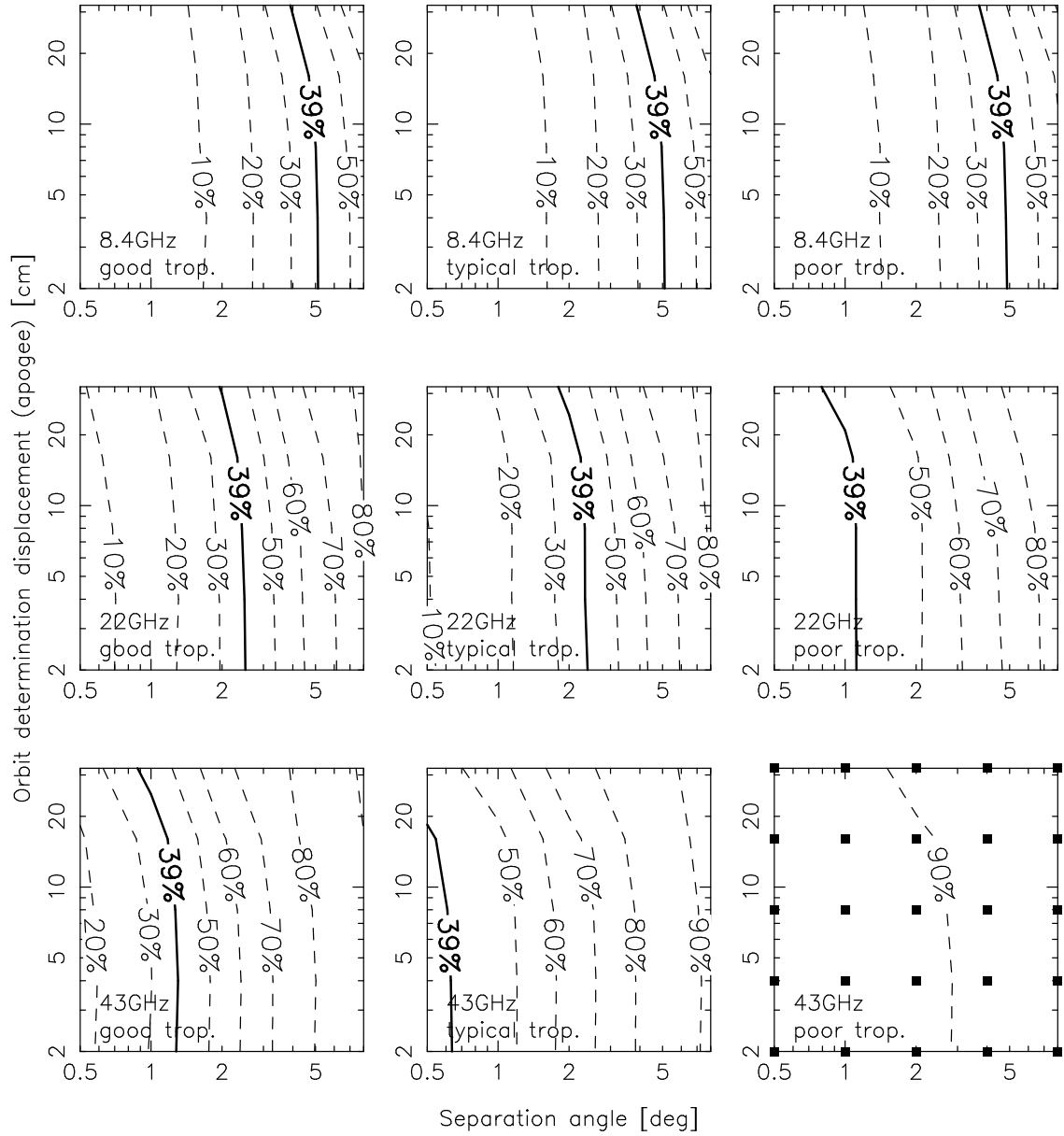

\begin{center}
\begin{tabular}{c}
    \FigureFile(150mm,150mm){figure-10.epsi}
\end{tabular}
\caption{
\label{fig:04-09}
Image coherence loss of the VSOP-2 phase referencing direct 
image with all the baselines in the case of the 60-s switching 
cycle. The abscissa and ordinate are the separation angle and 
OD displacement at the apogee in log scale, respectively. 
The contours show the percentage of the degradation relative 
to the model flux density of the target (10, 20, 30, 39, 50, 60, 
70, 80, 90\%). The top, middle, and 
bottom rows show the cases of 8.4-, 22-, and 43-GHz simulations. 
There are three plots in each row showing the good, typical, 
and poor tropospheric conditions. 
The filled squares in the right-bottom plot represent simulated grid. 
}
\end{center}
\end{figure}

The loss of the image coherence for the 60-s switching cycle is shown 
in figure~\ref{fig:04-09} for the various separation angles and ODDsA. 
In general, the image coherence loss increases as the observing 
frequency and/or the tropospheric phase fluctuations increase 
because it becomes more difficult to correctly connect calibrator 
fringe phases between the scans. 
The image coherence loss also increases as the separation 
angle and/or ODDA increase because residual long-term 
phase variations distort the synthesized images. 
According to 
\citet{TMS2001}, 
temporally averaged visibility 
degrades as $e^{-\sigma_{\phi}^2/2}$ where $\sigma_{\phi}^2$ 
is the variance of a random Gaussian phase noise. 
In this report we tentatively put a threshold criterion 
of successful phase referencing onto the image coherence 
of 61\% ($e^{-\sigma_{\phi}^2/2}\simeq 0.61$ when 
$\sigma_{\phi}=1$~radian); in other words, 
image coherence loss of 39\%. 
At 8.4~GHz, phase referencing is promising for all the 
tropospheric conditions, even with the ODDA of a few tens 
of centimeters. The simulation results indicate that the 
separation angle is recommended to be smaller than 
$4^\circ-5^\circ$. 
At 22~GHz the separation angle is recommended to be smaller than 
$1^\circ$ under poor tropospheric conditions, while phase referencing 
can be useful for a separation angle within 
$2^\circ-3^\circ$ under good and typical tropospheric conditions. 
There is little difference in the image coherence loss at 22~GHz 
between the good and typical tropospheric conditions. 
This indicates that the 60-s switching cycle is short enough to 
cancel the short-term phase fluctuations at 22~GHz 
under typical tropospheric conditions, or better, but not enough 
under poor tropospheric condition. At 43~GHz it is not 
recommended that the 
phase referencing observations are conducted under poor 
tropospheric condition because serious coherence loss is 
caused by tropospheric phase fluctuations. 
Under typical tropospheric conditions, phase referencing 
can be useful if the separation angle is smaller than $1^\circ$. 
Under good tropospheric conditions, the separation angle is 
recommended to be smaller than $1^\circ-2^\circ$. 

\onecolumn
\twocolumn
\onecolumn
\begin{figure}
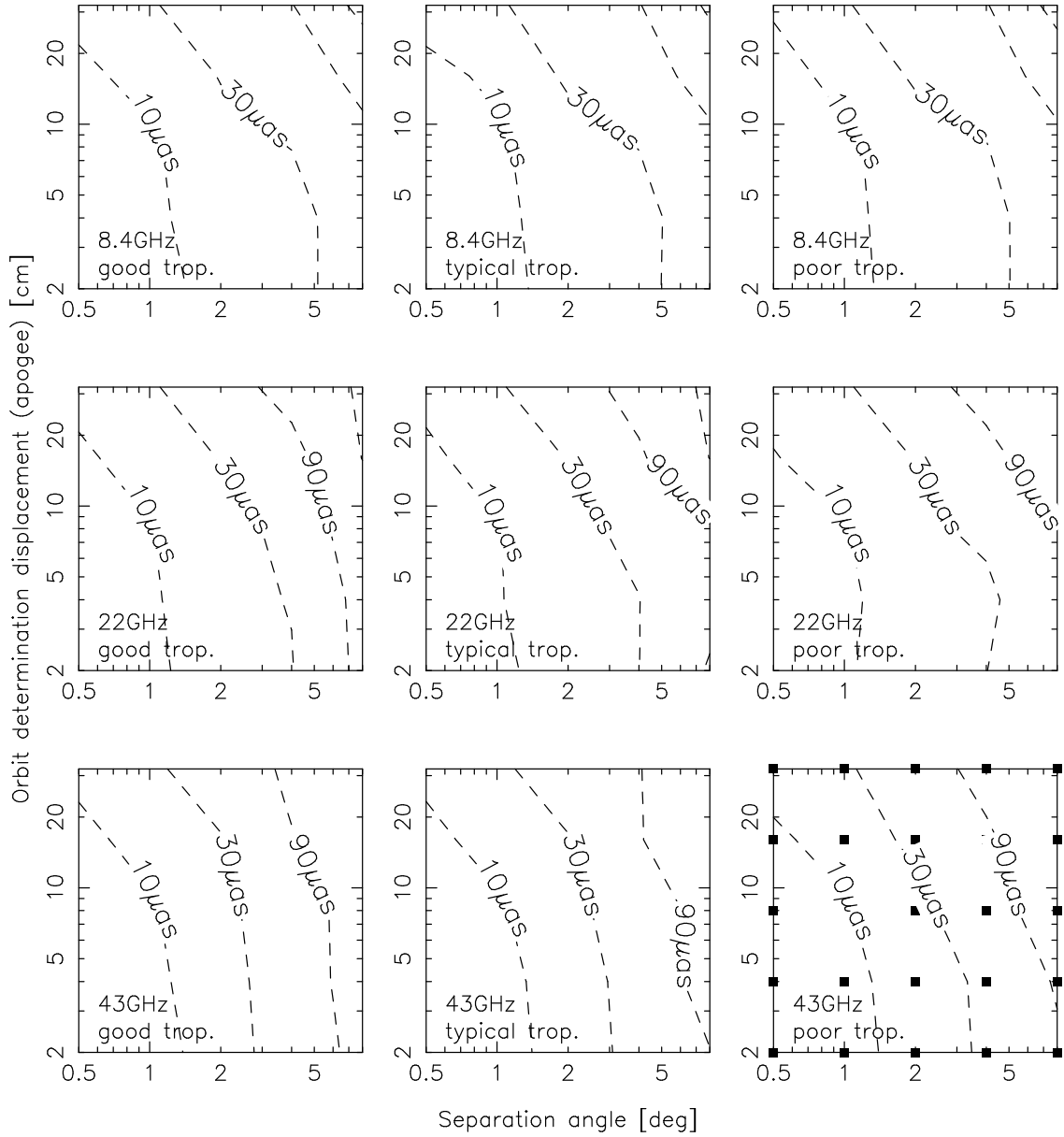

\begin{center}
\begin{tabular}{c}
    \FigureFile(150mm,150mm){figure-11.epsi}
\end{tabular}
\caption{
\label{fig:04-10}
Astrometric accuracy of of the VSOP-2 phase referencing direct image 
with all the baselines in the case of the 60-sec switching cycle. 
The contours show the positional offset 
of the image peak from the phase tracking center 
(10, 30, 90, 190~$\mu$as). The abscissa and ordinate are the 
separation angle and the OD displacement at the apogee in log scale, 
respectively. The top, middle, and bottom rows show the cases 
of 8.4-, 22-, and 43-GHz simulations. 
There are three plots in each row showing the good, typical, 
and poor tropospheric conditions. 
The filled squares in the right 
bottom plot represent simulated grid. 
}
\end{center}
\end{figure}
\twocolumn
\begin{figure}
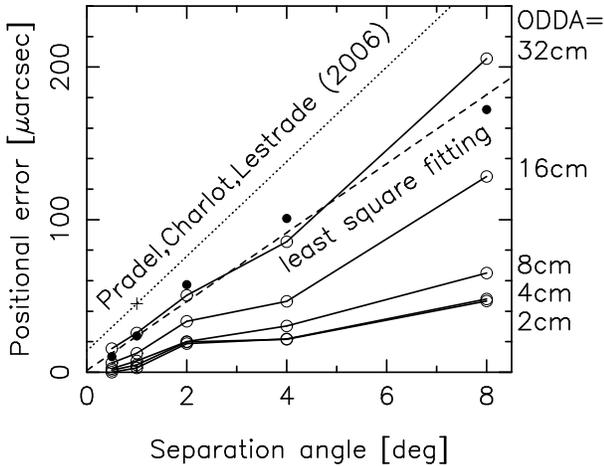

\begin{center}
\begin{tabular}{c}
    \FigureFile(80mm, 140mm){figure-12.epsi}
\end{tabular}
\caption{
\label{fig:04-11}
Astrometric accuracy of the surface brightness peak of the phase 
referencing direct image of the VLBA and VSOP-2. 
The abscissa is the separation angle, and the ordinate is the 
positional offset. 
Filled circles and the dashed line represent 
simulation results for the VLBA and 
the least square fitted line, respectively. 
Open circles represent VSOP-2 simulation results. 
The astrometric accuracy for the VLBA obtained with SPRINT 
software described by 
\citet{Pradel2006} 
is also shown by the cross mark and the dotted line 
with an assumption of their empirical formula. 
}
\end{center}
\end{figure}

The astrometric accuracy for the 60-s switching cycle are shown in 
figure~\ref{fig:04-10} for various separation angles and ODDsA. 
Phase referencing with VSOP-2 gives us an astrometric 
accuracy of better than 30~$\mu$as, in general, if the phase 
referencing direct image can be obtained without any 
serious image coherence loss. There are no prominent differences in the 
positional offset between the different tropospheric conditions 
as well as the different observing frequency bands. 
This is because the astrometric accuracy is directly related to 
the geometrical delay measurement accuracy. All of the VLBI delay errors, 
except for the ionospheric one, are non-dispersive for the observing 
frequency, and the ionospheric delay error is not serious at the VSOP-2 
observing frequency bands. Because the tendency about the 
astrometric accuracy is common for all of the simulation results 
in this study, we do not show the astrometric accuracy 
for the other cases mentioned below. 

At this stage let us consider possible advantages of VSOP-2 
phase referencing from the viewpoint of astrometry compared 
with terrestrial VLBI arrays. We carried out ARIS simulations 
of VLBA phase referencing observations for a comparison. 
We adopted the same parameters as given in 
tables~\ref{tbl:04-01} to \ref{tbl:04-04} for VLBA simulations, 
and the tropospheric condition was typical. 
The comparison described below was made at 8.4~GHz. 
In addition to the ARIS simulations, SPRINT software described by 
\citet{Pradel2006} 
(hereafter, PCL2006) was also used for VLBA simulations for 
$1^{\circ}$ pairs of sources with the center 
position at a declination of $59^{\circ}$ in order to check 
the behavior of ARIS. 
We carried out SPRINT simulations for 16 pairs of sources 
with the different position angles spaced equally. 
The phase referencing astrometric accuracy for a separation 
of $1^{\circ}$, 
$\Delta^{1^{\circ}}_{\alpha{\mathrm{cos}}\delta,\delta}$, 
defined by PCL2006, was obtained by averaging all of the pairs. 
Note that several of the parameters and conditions in the SPRINT 
simulations are different from those in ARIS: 
for example, the switching cycle time was 150 seconds, 
the lower elevation limit was $7^{\circ}$, 
the flux densities of both the target and the calibrator were 1~Jy, 
and not the ionospheric error, but the static component of 
the dry tropospheric error was considered. 
It should also be noted that the calibrator positional accuracy 
of 0.3~mas was set in SPRINT. Simulation results of the 
astrometric accuracy are shown in figure~\ref{fig:04-11} 
together with the simulation results of VSOP-2 phase referencing 
for the typical tropospheric condition, as already shown 
in figure~\ref{fig:04-10}. Assuming the empirical formula 
established by PCL2006 and the resultant 
$\Delta^{1^{\circ}}_{\alpha{\mathrm{cos}}\delta,\delta}$ 
of 45~$\mu$as from the SPRINT simulations, the ARIS results show 
systematically better performance than SPRINT. 
We carried out ARIS simulations for the case of 
the 150-s switching cycle; no distinguishable change could be 
seen from the 60-s switching cycle case. 
We then tested additional SPRINT simulations with an elevation 
limit of $20^{\circ}$ for a pair of sources 
separated along with the declination, and obtained 
$\Delta^{1^{\circ}}_{\alpha{\mathrm{cos}}\delta,\delta}$ 
of 38~$\mu$as, which is much more consistent with the ARIS 
simulation results. We consider that the discrepancy in the 
results between the two simulators might be attributed to 
the different elevation limit, and that this discrepancy 
can be much smaller if the simulation parameters are 
adjusted to match each other. 
We then compared the astrometric accuracy of the VLBA with VSOP-2. 
When the ODDA is less than several centimeters, the astrometric 
accuracy of VSOP-2 is much superior to that of the VLBA. 
PCL2006 indicated that the global VLBI array, which is the 
combination of the VLBA and European VLBI network, cannot improve 
the astrometric accuracy very much compared with the VLBA. 
In general, because the telescopes in the global VLBI array 
observe sources at low elevations for the longest baselines, 
the atmospheric systematic error must be more severe in 
those baselines. Since much longer baselines are available 
in VSOP-2 without serious atmospheric phase errors, VSOP-2 
phase referencing has an advantage compared with terrestrial VLBI 
arrays, unless the fringes of the space baselines 
are affected by the systematic phase error related to the OD. 
Assuming the tropospheric zenith error of 3~cm 
and the elevation limit of $20^{\circ}$, 
the allowable ODDA to achieve a higher astrometric accuracy with 
VSOP-2 can be roughly estimated at 
$3\times3~{\mathrm{cm}}/\sin{20^{\circ}}\sim26~{\mathrm{cm}}$, 
where the factor of three comes from the ratio between the maximum 
space and the terrestrial baseline lengths. 
From the simulation results, in cases where the ODDA is smaller 
than the 30-cm level, the astrometric accuracy of VSOP-2 is superior 
to the terrestrial VLBI arrays, as shown in figure~\ref{fig:04-11}. 

\begin{figure}
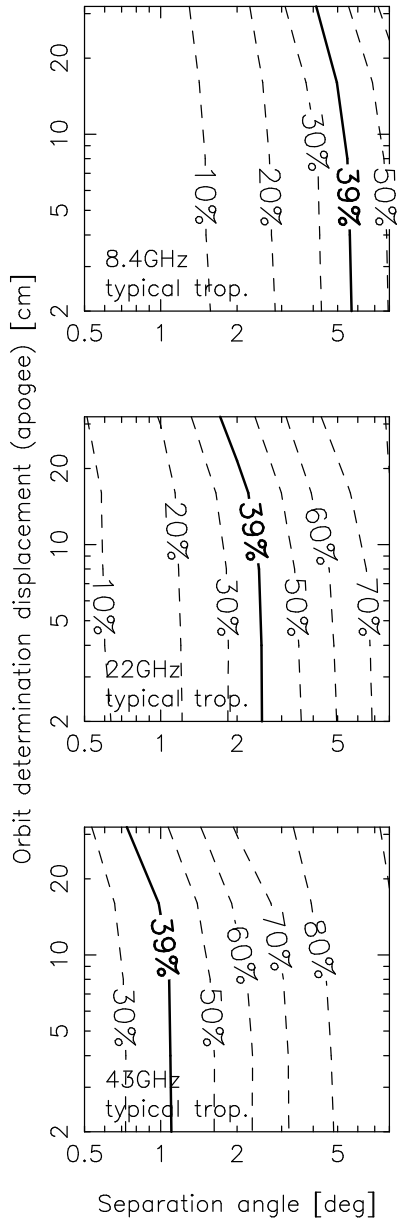

\begin{center}
\begin{tabular}{c}
    \FigureFile(51.5mm,150mm){figure-13.epsi}
\end{tabular}
\caption{
\label{fig:04-12}
Same as figure~\ref{fig:04-09} for the typical tropospheric condition, 
but the switching cycle time of 40~seconds. 
}
\end{center}
\end{figure}

For shorter switching cycle times, the image coherence 
can be improved for worse tropospheric conditions at the higher 
observing frequency bands. Figure~\ref{fig:04-12} shows the 
image coherence loss for the 40-s switching cycle under the typical 
tropospheric condition. 
At 43~GHz an improvement can clearly be seen in the 40-s switching cycle, 
while there is little difference between the switching cycle times of 
60 and 40 seconds at 8.4 and 22~GHz. We also conducted 
simulations for the 40-s switching cycle under good tropospheric 
conditions, and there was little improvement in all of the observing 
frequency bands. On the other hand, simulations at 22~GHz 
under poor tropospheric conditions showed an improvement, and 
the allowable separation angle was stretched to $2^\circ$. 
At 43~GHz it is not recommended that the phase referencing 
observations are conducted under poor tropospheric conditions, 
even with the 40-s switching cycle. 

\begin{figure}
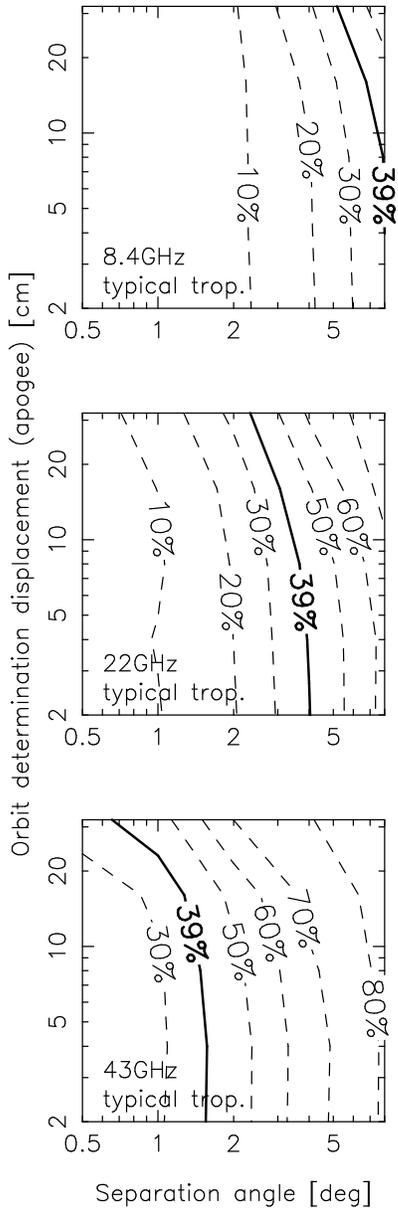

\begin{center}
\begin{tabular}{c}
    \FigureFile(51.5mm,150mm){figure-14.epsi}
\end{tabular}
\caption{
\label{fig:04-13}
Same as figure~\ref{fig:04-09} for the typical tropospheric 
condition, but the phase referencing direct images are made 
only with the space baselines. 
}
\end{center}
\end{figure}

The image coherence loss seems to be insensitive to the ODDA.  
This may be because there are only ten space baselines among 
55 baselines in the simulations. To investigate the influence 
of the ODDA on the image coherence, the phase referencing 
direct images were made only with 10 space baselines. 
The image coherence loss for the typical tropospheric condition 
and 60-s switching cycle time is shown in figure~\ref{fig:04-13}. 
The results show that the image coherence loss is more sensitive 
to the ODDA than that with all of the baselines, and is almost 
insensitive to an ODDA smaller than $\sim 10$~cm. 
This indicates that the 10-cm level ODDA meets VSOP-2 phase 
referencing. 
We note that the image coherence in figure~\ref{fig:04-13} is 
better compared with that in figure~\ref{fig:04-09}. This is because 
the fringe phase of the space baseline includes a single atmospheric 
phase error, while that of the terrestrial baseline includes a double one, 
so that the latter is less stable than the former, 
especially for small ODDsA. This does not mean, however,
that the phase referencing direct image only with the space 
baselines is superior to the image with all of the baselines, 
because the image RMS noise of the former is worse than the latter. 

The above recommendations may be too strict in some important 
applications. For example, 
the fringe at 43~GHz was subject to a serious coherence loss 
under poor tropospheric conditions, but an isolated peak could 
be seen in simulated images at the expected target position if 
the separation angle was rather less than $1^\circ$. Although 
the amplitude degradation cannot be recovered with the following 
self-calibration because the time scales of the residual phase 
fluctuations are typically shorter than a few minutes, 
the amplitude can be corrected by multiplying a factor of 
$e^{\sigma_{\phi}^{2}/2}$ for the baseline, where 
$\sigma_{\phi}$ is the RMS phase. Thus just for the astrometry, 
observers are less worried about the tropospheric conditions than 
those who are interested in the target morphology.

\begin{figure}
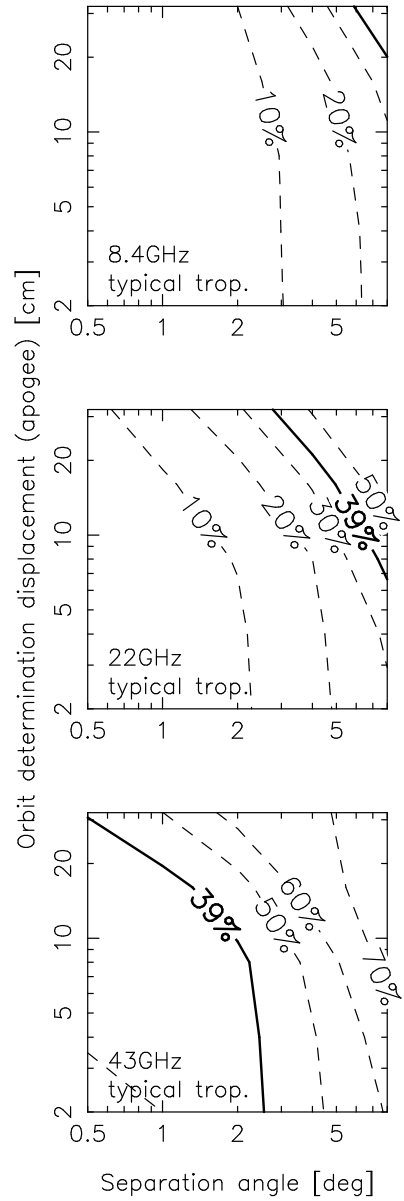

\begin{center}
\begin{tabular}{c}
    \FigureFile(51.5mm,150mm){figure-15.epsi}
\end{tabular}
\caption{
\label{fig:04-14}
Same as figure~\ref{fig:04-09} for the typical tropospheric 
condition, but with no atmospheric systematic errors. 
}
\end{center}
\end{figure}

In general, the simulation results show that the image coherence 
is insensitive to an ODDA smaller than $\sim 10$ cm. 
However, we obtained another indication of whether the atmospheric 
systematic errors were not included in the simulations. 
Figure~\ref{fig:04-14} shows the case with no atmospheric 
systematic errors under the typical tropospheric condition for 
the 60-s switching cycle phase referencing with all of the 
baselines. The performance was greatly improved comparing 
with the cases involving atmospheric systematic errors. 
In addition, the smaller the ODDA, the wider can the 
allowable separation angle be. 
\citet{Reid1999} showed that the atmospheric systematic 
errors should be calibrated for very precise astrometric 
VLBI observations at 43~GHz. They used a pair of sources to 
calibrate the atmospheric zenith delay errors by measuring 
the phase difference between them. 
Two phase referencing calibrators, or more, would be needed 
for the terrestrial array to calibrate 
the atmospheric systematic errors towards the best performance 
of VSOP-2 phase referencing. 

\clearpage

\subsection{
  Allowable Calibrator Flux Densities
}

It is plausible that the effectiveness of phase referencing 
decreases with fainter calibrators, 
because the fainter the calibrator, the more difficult 
will be the phase connection, due to thermal noise. 
We carried out another simulation series by varying the 
calibrator flux density from $10^{-5}$ to $10^{-4}$ of 
$\overline{S_{\mathrm{SEFD}}}$ of a space baseline 
with a separation angle of $1^\circ$ and an ODDA of 4~cm. 
In this simulation series, the calibrator flux densities 
ranged over $12-120$, $14-145$, and $23-233$~mJy at 8.4, 22, 
and 43~GHz, respectively, and the target flux densities were 
the same as listed in table \ref{tbl:04-04}. 

\begin{figure}
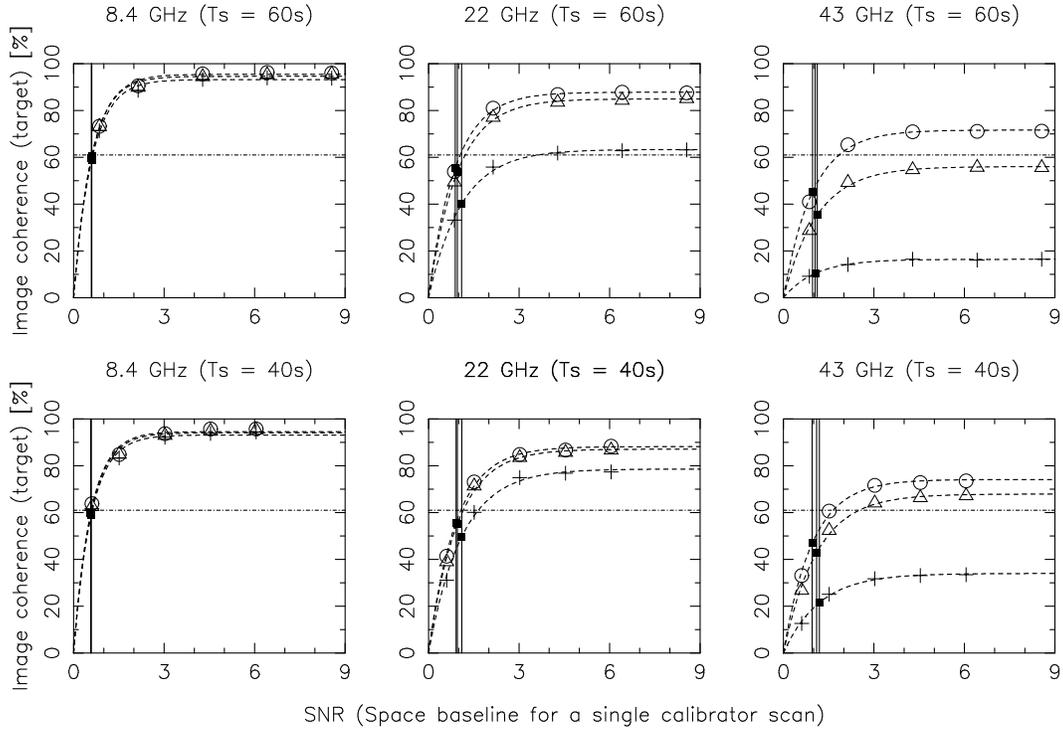

\begin{center}
\begin{tabular}{c}
    \FigureFile(140mm,140mm){figure-16a.epsi} \\
    \FigureFile(140mm,140mm){figure-16b.epsi}
\end{tabular}
\caption{
\label{fig:04-15}
Image coherence of the VSOP-2 phase referencing direct image 
with all the baselines as a function of signal-to-noise ratio 
of a space baseline for a single calibrator scan, $SNR^{\mathrm{c}}$. 
The top and bottom rows show the cases of 60-s and 40-s 
switching cycle times, respectively. All the simulation has the 
condition of $1^\circ$ separate pairs of sources and 4-cm ODDA. 
The calibrator flux densities in the simulations 
correspond to 10, 25, 50, 75, and 100\% of $10^{-4}$ of 
the geometric mean of SEFDs of the SRT and a VLBA telescope. 
The open circles, triangles, and crosses represent the good, 
typical, and poor tropospheric conditions, respectively. 
Horizontal dotted lines represent the crossed position to 
the threshold (39\% image coherence loss). The dashed lines 
are fitted curves to 
$Q_{\mathrm{0}} (1 - e^{-\xi \cdot SNR^{\mathrm{c}}})$ 
for each tropospheric condition. The filled squares and 
vertical lines represent the positions where 
$\xi \cdot SNR^{\mathrm{c}}=1$. 
}
\end{center}
\end{figure}

Figure \ref{fig:04-15} shows the image coherence obtained with 
all of the baselines as a function of the signal-to-noise ratio 
of a space baseline for a single calibrator scan. The image coherence 
drops as the calibrator flux density or the signal-to-noise ratio 
becomes smaller. The image coherence, $Q$, under the same tropospheric 
condition at the same observing frequency band can be well fitted to 
$Q
   = Q_{\mathrm{0}} (1-e^{-\xi \cdot SNR^{\mathrm{c}}})$, 
where $Q_{\mathrm{0}}$ and $\xi$ are the fitting parameters, and 
$SNR^{\mathrm{c}}$ is the signal-to-noise ratio of a space baseline 
for a single calibrator scan, represented by 
$
\eta (S^{\mathrm{c}} /~\overline{S_{\mathrm{SEFD}}})
  \sqrt{2\Delta\nu T^{\mathrm{c}}}
$.
It is noted from figure~\ref{fig:04-15} that the fitted 
curve shapes at 22 and 43~GHz resemble each other, 
while those at 8.4~GHz differ. This is because 
the ratios of 
$\overline{S_{\mathrm{SEFD}}}$ between the space and terrestrial 
baselines at 22 and 43~GHz are 2.2, by chance, while the ratio 
at 8.4~GHz is 15.3. Therefore, the image qualities at 8.4~GHz are 
more supported by the sensitivity of the terrestrial baselines than 
those at 22 and 43~GHz. 

\begin{figure}
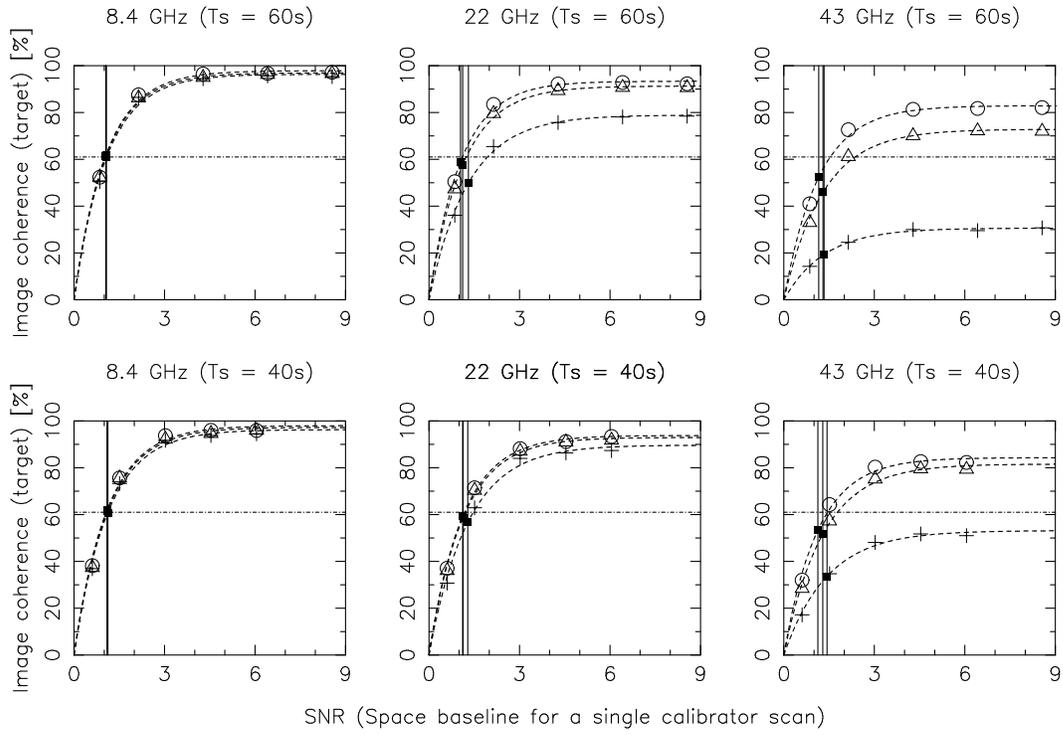

\begin{center}
\begin{tabular}{c}
    \FigureFile(140mm,140mm){figure-17a.epsi} \\
    \FigureFile(140mm,140mm){figure-17b.epsi}
\end{tabular}
\caption{
\label{fig:04-16}
Same as figure~\ref{fig:04-15}, but only with the space baselines. 
}
\end{center}
\end{figure}

Figure~\ref{fig:04-16} shows the simulation results of the 
image coherence with only the space baselines. It is noted 
that $\xi$ ranges over 1.1--1.4 under all the tropospheric 
conditions at all observing frequency bands. We carried out 
simulations with two other parameter combinations of 
the separation angle and ODDA 
($2^\circ$ separation and 4-cm ODDA, and $1^\circ$ separation 
and 8-cm ODDA), and found that there is no 
prominent difference in $\xi$ for the various parameter 
combinations. From simulations with only the space 
baselines, $\xi$ seems to be uniquely determined by the 
sensitivity of baselines, while $Q_{\mathrm{0}}$ is 
dependent on the observing frequency, tropospheric 
condition, switching cycle time, separation angle, and ODDA. 

The image coherence loss due to thermal noise is kept below 
2--4\% for an $SNR^{\mathrm{c}}$ of 4. Since the least sensitive 
baselines will be the space baselines in VSOP-2, it is safe for 
successful phase connection to choose calibrators with an 
$SNR^{\mathrm{c}}$ larger than 4. An $SNR^{\mathrm{c}}$ of 4, 
however, may lead to a false fringe detection in the fringe fitting 
for the calibrator. For good and typical tropospheric conditions, 
the coherence time at 22 and 43~GHz can be extended to be a few 
minutes, so that a possible solution to avoid false detection 
is to set the averaging time in the fringe fitting to a time 
interval including two to three calibrator scans. 

Another possible way is to make frequent fringe-finder scans 
interleaved in a nominal phase referencing observation: 
because a fringe-finder, such as an ICRF source, is 
bright and has a well-known position, fringe-finder scans are 
used to check the observing system and to calibrate the clock 
synchronization errors in VLBI observations. If the delay 
synchronization errors can be calibrated by group delay solutions 
with an accuracy of 1~ns, phase-slope in the single IF bandwidth 
of 128~MHz will cause only a 3\% coherence loss. This might lead 
to the conclusion that the phase referencing calibrators can 
be used without any fringe fitting before phase referencing, 
and that we may use the calibrators with an $SNR^{\mathrm{c}}$ 
of 4. To observe the fringe-finders frequently, a high-speed 
maneuvering capability for rather larger separation angles 
compared to the switching maneuvering will be needed. 

\onecolumn
\twocolumn
\onecolumn
\twocolumn
\section{
  Discussion
}

\subsection{
  Calibrator Source Counts
}

In the previous section we obtained constraints on the calibrator 
flux densities for the successful phase referencing with VSOP-2. 
For the VLBA as a terrestrial array, calibrators are recommended 
to have flux densities larger than 
56, 68, and 109~mJy at 8.4, 22, and 43~GHz, respectively, in the 
case of a 20-s calibrator scan, 
and 79, 96, and 154~mJy, respectively, in the case of a 10-s 
calibrator scan. Here we made use of two VLBI survey catalogues 
at $S$-band and $X$-band to investigate the probability to find 
a suitable calibrator located closely to an arbitrary target. 
The first is the VLBA calibrator catalogue, which can be 
accessed 
on-line\footnote{
$<$http://www.vlba.nrao.edu/astro/calib/vlbaCalib.txt$>$. 
}. 
This catalogue contains 3357 sources based on the VLBA Calibrator 
Survey (VCS) 
(\cite{Beasley2002}; 
\cite{Fomalont2003}; 
\cite{Petrov2005}; 
\cite{Petrov2006}). 
The ICRF sources are included in the catalogue as well. 
We can refer positions, integrated map flux densities, 
and unresolved map flux densities of compact sources 
observed with the VLBA, whose longest baseline is 
nearly 8000~km. The second is 
the Goddard Space Flight Center (GSFC) astrometric and 
geodetic catalogue, 
``2005f astro catalogue of compact radio sources 
(2005f\_astro)"\footnote{
$<$http://vlbi.gsfc.nasa.gov/solutions/2005f\_astro/$>$.
}. 
This catalogue provides calibrated visibility FITS-IDI data 
on the web as well as an html version of the source list. 
The catalogue 2005f\_astro contains 3481 objects of non-VCS 
sources observed for VLBI astrometry, and many of the VCS 
and ICRF sources included in the VLBA calibrator catalogue. 

We fitted the provided visibility amplitude $S(\nu, B)$ at 
each band to the following Gaussian function: 
\begin{eqnarray}
\label{equ:05-01}
S(\nu, B) &=& S_{0}(\nu) \exp{[-2(\pi \nu B \delta(\nu) / c)^2]},
\end{eqnarray}
where $S_{0}(\nu)$ and $\delta(\nu)$ are fitting parameters 
representing flux density with zero baseline and the size 
of the VLBI component, respectively. 
If the FITS-IDI data was found for a source in 2005f\_astro, 
the visibility amplitude versus projected baselines were 
fitted to equation~(\ref{equ:05-01}) to estimate 
$S_{0}$ and $\delta$. If it wasn't, we referred to 
the VLBA calibrator catalogue to use the integrated map flux 
density as $S_{0}$, and the unresolved map flux density as 
the visibility amplitude obtained with an 8000-km baseline. 
To extrapolate $S_{0}$ and $\delta$ at 22 and 43~GHz 
we assumed $\delta\propto\nu^{-\alpha}$ 
\citep{Cao2002} 
and $S_{0} \propto \nu^{-\beta}$, where $\alpha$ and $\beta$ 
were determined by fitting the $S$-band and the $X$-band results. 
We then inferred the visibility amplitude with a 25000-km 
projected baseline from equation~(\ref{equ:05-01}). 
If there were multi-epoch FITS-IDI data for a source 
in 2005f\_astro, we chose the largest value as the expected 
flux density at each VSOP-2 observing frequency band. 
If $S$-band data was not available for a source, only the 8.4-GHz 
estimate for a space baseline was made. If $X$-band data was not 
available for a source, none of the estimates was made. 

\begin{figure}
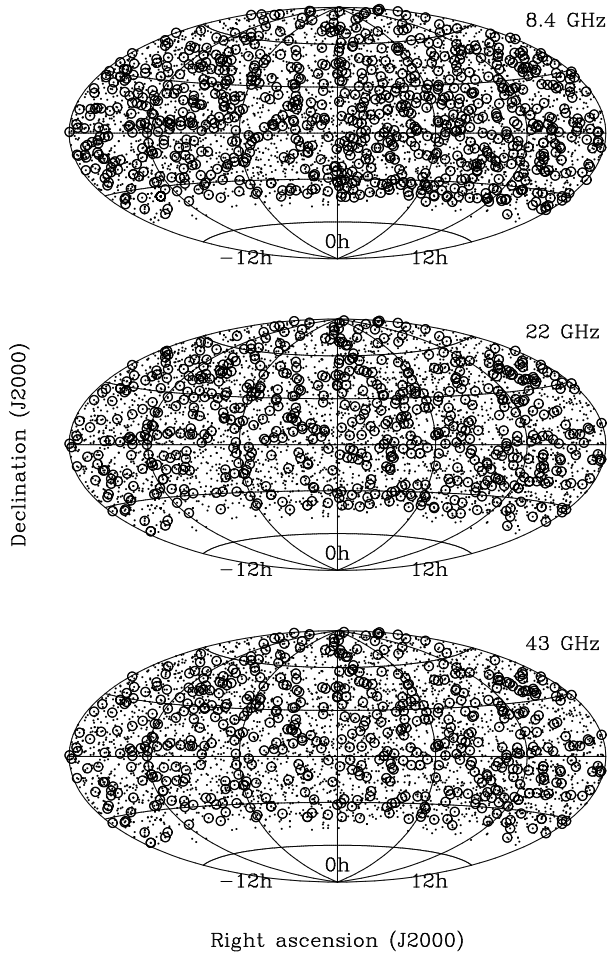

\begin{center}
\begin{tabular}{c}
    \FigureFile(80mm,80mm){figure-18.epsi}
\end{tabular}
\caption{
\label{fig:05-01}
Calibrator candidates for VSOP-2 phase referencing inferred from 
VLBI surveys at $S$-band and $X$-band. 
Dots represent sources used in the analysis. 
The open circles represent calibrator candidates 
with inferred flux densities for a projected baseline of 
25000~km larger than 50, 60, and 100~mJy at 8.4, 22, and 43~GHz, 
respectively.
}
\end{center}
\end{figure}

Figure~\ref{fig:05-01} shows the distribution of the calibrator 
candidates with inferred flux densities larger than 50, 60, and 
100~mJy at 8.4, 22, and 43~GHz, respectively. The number of 
candidates decreases as the observing frequency increase. 
Based on the calibrator candidate distribution, Monte Carlo 
simulations were performed to estimate the probability to find 
an adjacent phase referencing calibrator. The sky coverage for 
the Monte Carlo simulations was restricted to the northern sky 
because of the lack of the deep southern VLBI catalogue. 
The maximum separation angles to search for a calibrator 
were $5^\circ$, $3^\circ$, and $2^\circ$ 
at 8.4, 22, and 43~GHz, respectively. 
Our Monte Carlo simulations showed that the probabilities 
are 86, 43, and 20\% at 8.4, 22, and 43~GHz, respectively. 
The requirements for the calibrators and the simulation 
results are summarized in table \ref{tbl:05-01}. 
At 8.4~GHz, such a high probability is expected for VSOP-2 
phase referencing. The probabilities decrease as the 
frequency increases, because the requirements of the 
separation angle and flux density become greater. 
Since the Gaussian fitting of the visibility amplitude as a 
function of the projected baseline was conducted at the same 
frequency, 
we are confident that VSOP-2 phase referencing at 8.4~GHz 
is promising from the viewpoint of the calibrator availability. 
Observation chances with phase referencing at 22 and 
43~GHz, however, will be restricted in VSOP-2. On the 
other hand, the result at 22 and 43~GHz will be inspected 
more carefully because we adopted simple assumptions of 
$S_{0}$ and $\delta$ to estimate the probabilities. 

At 22 and 43~GHz, more sophisticated phase calibration schemes 
may be needed in addition to phase referencing to circumvent 
less calibrator availability in VSOP-2. 
One of the solutions is to observe with fainter calibrators. 
Assuming a terrestrial radio telescope with a 100-m diameter, 
calibrators with the flux densities of about 50~mJy will be 
available at 43~GHz. We conducted another Monte Carlo simulation 
at 43~GHz with the minimum calibrator flux density of 50~mJy 
and found that the probability was improved from 20\% to 25\%. 
However, because large telescopes can switch the sources much 
more slowly, it is not effective to use large telescopes in phase 
referencing at higher observing frequency bands. 
\citet{Doi2006} 
showed another possibility to use faint calibrators in 
phase referencing: they demonstrated their well-organized 
phase referencing observations, called bigradient phase referencing, 
using a combination of a very closely located faint calibrator and 
another bright but rather separate calibrator. 
Fringes of the target and faint calibrator are detected using 
the phase referencing technique with the bright calibrator. 
The long-term phase variations, because of the separation 
to the bright calibrator, are then calibrated with the closely 
located calibrators. Although their demonstrations were made 
at the $X$-band, the proposed method is promising for VSOP-2 
at the higher observing frequency bands. 
A phase-calibration technique with a water vapor radiometer 
(WVR) is also expected to use fainter calibrators in phase 
referencing. In this method a WVR is mounted on a terrestrial 
telescope to measure the amount of water vapor along the line 
of sight. An example of the successful application of this 
method is described by 
\citet{Marvel1998}. 
The WVR phase calibration technique will be able to achieve 
longer switching cycle time by removing the tropospheric phase 
fluctuations. This leads to longer scan durations for both 
target and calibrator in phase referencing observations, 
so that we can obtain a higher signal-to-noise ratio for faint 
calibrators. A longer switching cycle time provides further 
support to use large terrestrial telescopes with rather slow 
slew speeds. 

Another solution is to observe calibrators at lower frequency 
bands where a larger number of calibrator candidates is available. 
The refractivity of the water vapor is almost constant 
(non-dispersive) for radio waves. Thus, tropospheric phase 
fluctuations can be calibrated between different frequencies 
\citep{Asaki1998}. 
VLBI phase referencing experiments with different frequencies 
between two sources have been successfully demonstrated by 
\citet{Middelberg2005}. 
However, since the ionospheric excess path delay is dispersive, 
it is not possible to correct the ionospheric phase errors 
with calibrator phases at different frequencies 
\citep{Rioja2005}. 
Additional ARIS simulations were conducted for this multi-frequency 
phase referencing, and it was found that this method works 
for the observing frequency combination of 43 and 22~GHz for 
the target and calibrator, respectively. 
On the other hand, when calibrator's frequency is 8.4~GHz, 
large phase offset and fluctuated time variation remain in the 
compensated 22- or 43-GHz fringe phases due to the ionospheric 
excess path delays. We have to note that, during summer nights, 
much larger TEC disturbances than the 50-percentile amplitude 
are often observed, as indicated in figure \ref{fig:04-02}. 
It should also be noted that the influence of the ionosphere 
may be severer than expected above, because the solar activity 
is likely to reach its maximum in about 2012, when the VSOP-2 
satellite is planned to be launched. 
The above methods introduced here can improve the effectiveness 
of phase referencing under limited situations. It is important 
to carefully consider which method works most effectively for 
the target if some of them are available. 

\subsection{
  Satellite Attitude Control
}

Phase referencing has been quite successful when a pair of 
sources are so close that both of them are observed in the 
same beam of each radio telescope. In such very fortuitous 
cases the formal errors of the relative position measurements 
are as low as several $\mu$as 
(e.g., \cite{Marcaide1983}; 
\cite{Sudou2003}). 
Since HALCA does not have an ability to change the antenna pointing 
so quickly, a so-called in-beam phase referencing has been 
carried out in VSOP
(\cite{Bartel2000}; 
\cite{Porcas2000}; 
\cite{Guirado2001}). 

A technical challenge in the attitude control of the satellite 
will be made in VSOP-2 to provide a powerful solution for 
regularly switching maneuvering. For phase referencing at 43~GHz 
the satellite is required to repeatedly maneuver between two 
celestial sources separated by a few degrees every a few tens 
of seconds, and observe sources with an attitude stability 
of $0.005^{\circ}$. Since such a switching maneuvering is 
hardly achievable with standard satellite reaction wheels (RWs), 
two control moment gyros (CMGs) with a single-gimbaled flywheel 
spinning at a constant angular rate are planned to be added for 
fast attitude switching. A CMG is a momentum-exchange device 
that can produce large output torque on the satellite body 
by rotating the gimbal axes. 
The switching maneuvers around the two orthogonal axes 
for a pair of sources (roll and pitch) are made by controlling 
the torques provided by the two CMGs while four RWs mainly 
control the attitude around the axis to the sources (yaw) 
by generating control torques. The switching maneuvers during 
a phase referencing observation are the round trip between 
a pair of sources, so that the two CMGs generate the torques 
to switch back and forth with no net change in the total 
angular momentum of the CMG system by an operational symmetry. 
Numerical simulations for the rigid body with the CMGs showed 
that a $3^{\circ}$ switching maneuver within 15~seconds 
and tracking the sources during a scan is possible for 
the 60-s switching cycle time together with a wide range gyroscope. 
An antenna dynamical model has also been developed in a 
feasibility study of the switching maneuvering because the 
deployable main reflector and boom connecting the reflector 
and satellite body will be major sources to excite 
eigen-frequencies causing the attitude disturbance. The satellite 
will be designed so as not to excite the eigen-frequencies lower 
than 0.25~Hz for stabilizing the pointing between the switching 
maneuvers. 

A high-speed maneuvering capability with the RWs 
is important to increase the efficiency of the operation 
of the space VLBI. The capability of the attitude 
maneuvering over large angles with a rate of $0^{\circ}.1$~s$^{-1}$ 
will also be useful for fringe-finder scans interleaved 
in a phase referencing observation. 

\subsection{
  Highly Accurate Orbit Determination
  \label{OD_METHOD}
}

A highly accurate OD, with a positional accuracy of better than 
several centimeters, is required for VSOP-2 phase referencing. 
This requirement is two orders of magnitude better than the OD 
accuracy achieved for HALCA by Doppler tracking. One of the possible 
methods to achieve the OD required for VSOP-2 phase referencing 
is to use the on-board GPS receiver: by using on-board GPS 
navigation systems, the TOPEX/POSEIDON satellite launched 
in 1992 to measure the ocean surface level achieved the OD accuracy 
of 2--3\,cm, and GRACE, to measure the center of the gravity of 
the Earth, achieved an OD accuracy of about 1--2~cm. These 
satellites are in relatively low-Earth orbits, so that more 
than several GPS satellites, whose altitudes are $\sim20200$~km, 
are always available. When the user altitude is higher than 
$\sim3500$~km, it will be outside of the main beam illumination 
of a given GPS satellite, because the beam width of the GPS 
transmitting antenna is designed to illuminate near-Earth users. 
At altitudes near the VSOP-2 satellite apogee, zero to only three 
GPS satellites can be available at any given time, even with the 
on-board GPS receiver antenna system covering all directions. 

To strengthen the orbit determination, high-quality accelerometers 
can be used together with the GPS measurements. Accelerometry 
will connect the orbit positions and the velocities over 
periods of time when the GPS measurements are unable to provide 
good solutions.
\citet{Wu2001} 
showed in their covariance analysis for the VSOP-2 OD 
that the conventional GPS navigation with accelerometry 
can achieve an orbit formal error well below 1\,cm for 
the VSOP-2 satellite in all the three components near 
perigee. At higher altitudes, however, the OD error 
grows to about 2\,cm with the assumption of an on-board 
accelerometer of 1~nm~s$^{-2}$ accuracy due to the lack 
of GPS measurements at these altitudes.  

Another possibility is to have a GPS-like signal transmitter 
on the VSOP-2 satellite. It is expected that, in conjunction 
with the on-board GPS receiver, the OD accuracy will then be 
less than 1.5\,cm. To further improve the OD accuracy to the 
1-cm level, ultra-precise accelerometers at the level of 
0.1~nm~s$^{-2}$ now available should be used. It should be noted 
that missions like GRACE were carefully designed to have the 
accelerometer at the center of mass of the satellite, which may 
be difficult to achieve for the VSOP-2 satellite. To assure 
the 1-cm accuracy at all times, better determination of GPS 
orbits and clocks will be required. 

Galileo is a GPS-like navigation system planned in Europe, 
to be fully operational in 2008. The constellation consists 
of 30 satellites whose orbits are the circular with the 
altitude of 23616~km. Galileo satellites will be equipped with 
the hydrogen maser time standards and are expected to achieve 
more precise OD. This system is more effective for VSOP-2 and 
the 2.5-cm level OD can be achieved only with the use of 
GPS/Galileo receivers on the VSOP-2 satellite.

\begin{table}
\caption{
Probabilities to find a suitable calibrator for VSOP-2 phase
referencing.
}\label{tbl:05-01}
\begin{center}
\begin{tabular}{p{17mm}p{16mm}p{16mm}p{16mm}}
\hline
\hline

   & 8.4~GHz
   & 22~GHz
   & 43~GHz     \\
\hline
$S^{\mathrm{c}}$
  & $\geq$50~mJy
  & $\geq$60~mJy
  & $\geq$100~mJy \\
$\Delta\theta$
  & $\leq5^\circ$
  & $\leq3^\circ$
  & $\leq2^\circ$ \\
\hline
Probability\footnotemark[$*$]
  & 86\%
  & 43\%
  & 20\% \\
\hline
  \multicolumn{3}{@{}l@{}}{\hbox to 0pt{\parbox{120mm}{\footnotesize
      \par\noindent
      \footnotemark[$*$]
        Monte Carlo simulations were performed for the northern sky.
    }\hss}}
\end{tabular}
\end{center}
\end{table}
\section{
  Conclusions
}

The effectiveness of phase referencing with VSOP-2 was verified 
in detail with a newly developed software simulation tool, ARIS. 
Simulations with ARIS show that phase referencing with VSOP-2 is 
promising at 8.4~GHz for all of the tropospheric conditions, 
while at 22 and 43~GHz the phase referencing observations are 
recommended to be conducted under good and typical tropospheric 
conditions. At 22 and 43~GHz there is another difficulty in 
terms of the calibrator choice: our ARIS simulations show that 
it is safe to choose a phase referencing calibrator with the 
expected signal-to-noise ratio on a space baseline larger than 4 
for a single calibrator scan, but such a bright calibrator cannot 
always be found closely enough to a given target at 22 and 43~GHz. 

The specification requirements of the satellite in terms of the 
maneuvering capability and OD were obtained from our 
investigations. At 22 and 43~GHz, one-minute or shorter 
switching capability is required, while a few minute or longer 
switching cycle times may be used at 8.4~GHz. An accuracy of 
the orbit determination of less than $\sim10$~cm is 
required for the mission. Current studies concerning the VSOP-2 
satellite design indicate prospects that it is not easy, but 
not impossible, to achieve. 

Although the atmospheric systematic errors cannot perfectly be 
removed with the a priori values calculated in the correlator, 
those phase errors can be corrected in well-organized phase referencing 
observations along with multiple calibrators. Note that the satellite 
does not need to observe multiple calibrators in a short period, 
because the systematic errors are related to the terrestrial telescopes. 
If the atmospheric systematic errors can be successfully removed, 
a few centimeter OD accuracy will be targeted so that the performance 
of VSOP-2 phase referencing will be greatly improved. 

In this report we have demonstrated the usefulness of 
ARIS in investigating the effectiveness of VSOP-2 phase referencing. 
ARIS will also be convenient to check VLBI observation plans 
from the viewpoint of image quality. In this report we considered 
some of the intended cases of VSOP-2 phase referencing observations 
for point sources with the highest spatial resolution; further 
investigation can be made for an individual source. This is 
important for the VSOP-2 scientific goals, especially at 22 and 
43~GHz, because the phase referencing technique cannot always 
be used at those frequency bands in terms of finding calibrators. 
ARIS will give helpful suggestions to comprise the effective 
observation and operation plans for the best performance in VSOP-2. 

The authors made use of the GPS TEC data taken by GSI and 
provided by Kyoto University. The authors made use of 
the VLBA calibrator catalogue of NRAO and 2005f\_astro 
catalogue of NASA GFSC. The authors express their hearty 
thanks to all members of VSOP-2 project team, especially, 
H.~Hirabayashi of ISAS who is leading the next space VLBI 
working group and M.~Inoue of NAOJ space VLBI project office. 
The authors also express their thanks to 
S-C.~Wu and Y.~Bar-Sever of the Jet Propulsion Laboratory 
for their investigations of the VSOP-2 satellite OD. 
Y.~Asaki gives his thanks to T.~Ichikawa of ISAS for 
useful suggestions about the satellite OD, K.~Noguchi 
of Nara women's university for discussion about the 
ionospheric TEC fluctuations, and L.~Petrov of NASA GSFC 
for discussion about the VLBI compact radio source surveys, 
and D.~Jauncey for comments about this work.


\begin{thebibliography}{}
\bibitem[Asaki et al.(1996)]{Asaki1996} 
    Asaki,~Y., Saito,~M., Kawabe,~R., Morita,~K-I., \& Sasao,~T. 
    1996,
    Radio~Sci., 
    31,
    1615 
\bibitem[Asaki et al.(1998)]{Asaki1998}
    Asaki,~Y., Shibata,~K.~M., Kawabe,~R., Roh,~D-G., Saito,~M., Morita,~K-I., 
    \& Sasao,~T.  
    1998,
    Radio~Sci., 
    33,
    1297
\bibitem[Bartel et al.(1986)]{Bartel1986}
    Bartel,~N., Herring,~T.~A., Ratner,~M.~I., Shapiro,~I.~I., \& Corey,~E. 
    1986,
    \nat, 
    319,
    733
\bibitem[Bartel, Bietenholz(2000)]{Bartel2000}
    Bartel,~N., \& Bietenholz,~M.~F. 
    2000,
    in Astrophysical Phenomena Revealed by Space VLBI, 
    ed.\ H.~Hirabayashi, P.~G.~Edwards, \& D.~W.~Murphy (ISAS, Sagamihara), 
    17
\bibitem[Beasley, Conway(1995)]{Beasley1995}
    Beasley,~A.~J., \& Conway,~J.~E. 
    1995, 
    in Very Long Baseline Interferometry and the VLBA, 
    ed.\ J.~A.~Zensus, P.~J.~Diamond, \& P.~J.~Napier 
    (ASP~Conf.~Ser., 82), 
    327
\bibitem[Beasley et al.(2002)]{Beasley2002}
    Beasley,~A.~J., Gordon,~D., Peck,~A.~B., Petrov,~L., 
    MacMillan,~D.~S., Fomalont,~E.~B., \& Ma,~C.
    2002, 
    \apjs, 
    141, 
    13
\bibitem[Bristow, Greenwald(1997)]{Bristow1997}
    Bristow,~W.~A., \& Greenwald,~R.~A. 
    1997,
    \jgr,
    102, 
    11585
\bibitem[Brunthaler et al.(2005)]{Brunthaler2005}
    Brunthaler,~A., Reid,~M.~J., Falcke,~H., Greenhill,~L.~J., \& Henkel,~C.
    2005,
    Science, 
    307, 
    1440
\bibitem[Carilli, Holdaway(1999)]{Carilli1999}
    Carilli,~C.~L., \& Holdaway,~M.~A. 
    1999, 
    Radio~Sci., 
    34, 
    817
\bibitem[Cao and Jiang(2002)]{Cao2002}
    Cao,~X, \& Jiang.,~D.~R.,
    2002, 
    MNRAS,
    331, 
    111
\bibitem[Cotton(1995)]{Cotton1995}
    Cotton,~W.~D. 
    1995, 
    in Very Long Baseline Interferometry and the VLBA, 
    ed.\ J.~A.~Zensus, P.~J.~Diamond, \& P.~J.~Napier 
    (ASP~Conf.~Ser., 82), 
    189
\bibitem[Doi et al.(2006)]{Doi2006}
    Doi,~A., et al.
    2006, 
    \pasj, 
    58, 
    777
\bibitem[Dravskikh, Finkelstein(1979)]{Dravskikh1979}
    Dravskikh,~A.~F., \& Finkelstein,~A.~M.
    1979, 
    \apss, 
    60, 
    251
\bibitem[Fey et al.(2004)]{Fey2004}
    Fey,~A.~L., et al.
    2004, 
    \aj, 
    127, 
    3587
\bibitem[Fomalont(1995)]{Fomalont1995}
    Fomalont,~E. 
    1995, 
    in Very Long Baseline Interferometry and the VLBA, 
    ed.\ J.~A.~Zensus, P.~J.~Diamond, \& P.~J.~Napier 
    (ASP~Conf.~Ser., 82), 
    363
\bibitem[Fomalont et al.(2003)]{Fomalont2003}
    Fomalont,~E.~B., Petrov,~L., MacMillan,~D.~S., Gordon,~D., \& Ma,~C.
    2003, 
    \aj, 
    126, 
    2562
\bibitem[Gallimore, Beswick(2004)]{Gallimore2004a}
    Gallimore,~J.~F., \& Beswick,~R. 
    2004,
    \aj,
    127,
    239
\bibitem[Gallimore et al.(2004)]{Gallimore2004b}
    Gallimore,~J.~F., Baum,~S.~A., \& O'Dea,~C.~P. 
    2004,
    \apj,
    613, 
    794
\bibitem[Georges(1968)]{Georges1968}
    Georges,~T.~M. 
    1968,
    Journal of Atmospheric and Terrestrial Physics, 
    30, 
    735
\bibitem[Guirado et al.(2001)]{Guirado2001}
    Guirado,~J.~G., Ros,~E., Jones,~D.~L., Lestrade,~J.~-F., 
    Marcaide~J.~M., P{\'e}rez-Torres,~M.~A., \& Preston,~R.~A. 
    2001,
    \aap, 
    371,
    766
\bibitem[Gwinn et al.(1986)]{Gwinn1986}
    Gwinn,~C.~R., Taylor,~J.~H., Weisberg,~J.~M., \& Rawley,~R.~A. 
    1986, 
    \aj, 
    91, 
    338
\bibitem[Hirabayashi et al.(1998)]{Hirabayashi1998}
    Hirabayashi,~H., et al. 
    1998, 
    Science, 
    281, 
    1825
\bibitem[Hirabayashi et al.(2000)]{Hirabayashi2000}
    Hirabayashi,~H., et al. 
    2000, 
    \pasj, 
    52, 
    955
\bibitem[Hirabayashi et al.(2004)]{Hirabayashi2004}
    Hirabayashi,~H., et al. 
    2004, 
    \procspie, 
    5487, 
    1646
\bibitem[Ho et al.(1997)]{Ho1997}
    Ho,~C.~M., Wilson,~B.~D., Mannucci,~A.~J., Lindqwister,~U.~J., \& 
    Yuan,~D.~N. 
    1997, 
    Radio~Sci., 
    32, 
    1499
\bibitem[Lestrade et al.(1990)]{Lestrade1990}
    Lestrade,~J.~-F., Rogers,~A.~E.~E., Whitney,~A.~R., 
    Niell,~A.~E., Phillips,~R.~B~., \& Preston, R., 
    1990, 
    \aj, 
    99, 
    1663
\bibitem[Ma et al.(1998)]{Ma1998}
    Ma,~C., et al. 
    1998, 
    \aj, 
    116, 
    516
\bibitem[Marcaide, Shapiro(1983)]{Marcaide1983}
    Marcaide,~J.~M., \& Shapiro,~I.~I. 
    1983,
    \aj, 
    88, 
    1133
\bibitem[Marvel, Woody(1998)]{Marvel1998}
    Marvel,~K.~B., \& Woody,~D.~P.
    1998, 
    \procspie, 
    3357, 
    442
\bibitem[McCarthy, Petit(2004)]{McCarthy2004}
    McCarthy,~D.~D., \& Petit,~G.\ (ed.) 
    2004,
    IERS Technical Notes, 
    32,
    ch.5
\bibitem[Middelberg et al.(2005)]{Middelberg2005}
    Middelberg,~E., et al. 
    2005, 
    \aap, 
    433, 
    897
\bibitem[Migenes et al.(1999)]{Migenes1999}
    Migenes,~V., et al. 
    1999, 
    \apjs,
    123, 
    487
\bibitem[Murphy et al.(2005)]{Murphy2005}
    Murphy,~D., Preston,~R., Fomalont,~E., Romney,~J., Ulvestad,~J., 
    Greenhill,~L., \& Reid,~M. 
    2005, 
    in Future Directions in High Resolution Astronomy, 
    ed.\ J.~D.~Romney, \& M.~J.~Reid 
    (ASP~Conf.~Ser., 340), 
    575
\bibitem[Napier(1995)]{Napier1995}
    Napier,~P.~J. 
    1995, 
    in Very Long Baseline Interferometry and the VLBA, 
    ed.\ J.~A.~Zensus, P.~J.~Diamond, \& P.~J.~Napier 
    (ASP~Conf.~Ser., 82), 
    59
\bibitem[Niell(1996)]{Niell1996}
    Niell,~A.~E. 
    1996, 
    \jgr, 
    101, 
    3227
\bibitem[Noguchi et al.(2001)]{Noguchi2001}
    Noguchi.,~K., Imamura,~T., Oyama,~K.-I., \& Saito,~A. 
    2001, 
    Radio Sci.,
    36, 
    1607
\bibitem[Petrov et al.(2005)]{Petrov2005}
    Petrov,~L., Kovalev,~Y.~Y., Fomalont,~E., Gordon,~D. 
    2005, 
    \aj, 
    129, 
    1163
\bibitem[Petrov et al.(2006)]{Petrov2006}
    Petrov,~L., Kovalev,~Y.~Y., Fomalont,~E.~B., Gordon,~D. 
    2006, 
    \aj, 
    131, 
    1872
\bibitem[Porcas et al.(2000)]{Porcas2000}
    Porcas,~R.~W., Rioja,~M.~J., Machalski,~J., \& Hirabayashi,~H. 
    2000,
    in Astrophysical Phenomena Revealed by Space VLBI, 
    ed.\ H.~Hirabayashi, P.~G.~Edwards, \& D.~W.~Murphy (ISAS, Sagamihara), 
    245
\bibitem[Pradel, Charlot, Lestrade(2006)]{Pradel2006}
    Pradel,~N., Charlot,~P., \& Lestrade,~J.-F.
    2006,
    \aap,
    452,
    1099
\bibitem[Reid, Readhead, Vermulen, Treuhaft(1999)]{Reid1999}
    Reid,~M.~J., Readhead,~A.~C.~S., Vermeulen,~R.~C., \& Treuhaft,~R.~N. 
    1999, 
    \apj, 
    524, 
    816
\bibitem[Rioja et al.(2005)]{Rioja2005}
    Rioja,~M., Dodson,~R., Porcas,~R., Suda,~H., \& Colomer,~F. 
    2005, 
    in Proceedings of the 17th working meeting on European VLBI for 
    geodesy and astrometry, 
    ed.\ M.~Vennebusch, \& A.~Nothnagel 
    (INAF Istituto di Radioastronomia Sezione di NOTO, Italy), 
    125
\bibitem[Ros et al.(2000)]{Ros2000}
    Ros,~E., Marcaide,~J.~M., Guirado,~J.~C., Sard\'on,~E., \& Shapiro,~I.~I. 
    2000, 
    \aap, 
    356, 
    357
\bibitem[Saito et al.(1998)]{Saito1998}
    Saito.,~A, Fukao,~S., \& Miyazaki,~S. 
    1998, 
    \grl, 
    25, 
    3079
\bibitem[Shapiro et al.(1979)]{Shapiro1979}
    Shapiro,~I.~I., et al. 
    1979, 
    \aj, 
    84, 
    1459
\bibitem[Smith et al.(2003)]{Smith2003}
    Smith,~K, Pestalozzi,~M., G\"udel,~M., Conway,~J., \& Benz,~A.~O. 
    2003, 
    \aap, 
    406, 
    957
\bibitem[Sudou et al.(2003)]{Sudou2003}
    Sudou,~H., Iguchi,~S., Murata,~Y., \& Taniguchi,~Y. 
    2003, 
    Science, 
    300, 
    1263
\bibitem[Tatarskii(1961)]{Tatarskii1961}
    Tatarskii,~V.~I. 
    1961, 
    Wave Propagation in a Turbulent Medium (Dover, New York),
    ch.1
\bibitem[Thompson, Moran, Swenson(2001)]{TMS2001}
    Thompson,~A.~R., Moran,~J.~M., \& Swenson, Jr.,~G.~W. 
    2001, 
    Interferometry and synthesis in radio astronomy 
    (A Wiley-Interscience Publication, John Wiley \& Sons, Inc., New York), 
    ch.13
\bibitem[Treuhaft, Lanyi(1987)]{Treuhaft1987}
    Treuhaft,~R.~N., \& Lanyi,~G.~E. 
    1987,
    Radio~Sci., 
    22, 
    251
\bibitem[Wu, Bar-Sever(2001)]{Wu2001}
    Wu,~S-C., \& Bar-Sever,~Y. 
    2001,
    Proc. ION GPS 2001 (Salt Lake City, Utah), 
    2272
\end{thebibliography}
\end{document}